\documentclass[twocolumn,english,aps,prb,superscriptaddress, floatfix]{revtex4-2}

\usepackage[T1]{fontenc}
\usepackage{graphicx}
\usepackage{mathtools}
\usepackage{amsmath}
\usepackage{dsfont}
\usepackage{amsfonts}
\usepackage{amssymb}

\usepackage{hyperref}

\usepackage{color}
\usepackage{xcolor}
\usepackage{xspace}

\usepackage{braket} 
\usepackage{tikz}

\newcommand{\R}{\ensuremath{\mathbb{R}}\xspace}
\newcommand{\HV}{\ensuremath{\operatorname{HV}}\xspace}

\begin{document}

\title{Variational Quantum Multi-Objective Optimization}

\author{Linus Ekstr\o m}
\affiliation{Honda Research Institute Europe GmbH, Carl-Legien-Str.\ 30, 63073 Offenbach, Germany}
\author{Hao Wang}
\affiliation{LIACS/aQa, Leiden University, Niels Bohrweg 1, 2333 CA Leiden, Netherlands}
\author{Sebastian Schmitt}
\affiliation{Honda Research Institute Europe GmbH, Carl-Legien-Str.\ 30, 63073 Offenbach, Germany}
\date{February 5, 2024}

\begin{abstract}
Solving combinatorial optimization problems on near-term quantum devices has gained a lot of attraction in recent years. 
Currently, most works have focused on single-objective problems, whereas  
many real-world applications need to consider multiple, mostly conflicting objectives, such as cost and quality.
We present a variational quantum optimization algorithm
to solve discrete multi-objective optimization problems on quantum computers. 
The proposed quantum multi-objective optimization (QMOO) algorithm incorporates all cost Hamiltonians representing the classical objective functions in the quantum circuit and produces a quantum state consisting of Pareto-optimal solutions in superposition.
From this state we retrieve a set of solutions  and utilize the widely applied hypervolume indicator to determine its quality as an approximation to the Pareto-front. 
The variational parameters of the QMOO circuit are tuned by maximizing the hypervolume indicator in a quantum-classical hybrid fashion. 
We show the effectiveness of the proposed algorithm on several benchmark problems with up to five objectives. 
We investigate the influence of the classical optimizer, the circuit depth and compare to results from classical optimization algorithms. 
We find that the algorithm is robust to shot noise and produces good results with as low as 128 measurement shots in each iteration. 
These promising result open the perspective to run the algorithm on near-term quantum hardware.  
\end{abstract}

\maketitle

\section{Introduction}
Over the past few years, quantum computing experiments have indicated that the threshold of practically useful quantum computation is nearing~\cite{acharya2022suppressing, Youngseok2023evidence,queraLogical2023,quantinuumMicrosoft2024}.
Realistic application problems have increasingly moved into the focus of quantum algorithm research, in particular for the currently available noisy-intermediate scale quantum (NISQ) devices~\cite{NISQAlgo, deller_quantum_2023, yarkoni2021, abbas_QuantumOptmization_2023,tq_EVcharging2023,dalyacSmartChargingQAOA2021,tq_EVcharging2023}.
A promising class of approaches are the variational algorithms such as the quantum approximate optimization algorithm (QAOA)~\cite{farhi2014quantum,blekosQAOA_Review}. 
However, the exact nature of the potential quantum advantage for real-world combinatorial optimization problems remains an open problem~\cite{blekosQAOA_Review,farhi2015quantum, farhi2019quantum}. 

Current versions of variational quantum optimization approaches only deal with single-objective optimization problems.
The final target for the optimized quantum circuit is to produce a quantum state where the probability of sampling the optimal solution of the original problem is as large as possible. 
However, many realistic application problems need to consider more than one objective, such as cost and quality, simultaneously~\cite{SharmaMOOreview2022,pereiraMOOreview2022}.
Since these objectives are mostly in a trade-off relationship where increasing the performance in one objective leads to a degradation in the other objectives, finding one single optimal solution is no longer a proper target setting. 
Instead, the goal is to find a set of Pareto-optimal solutions that represent the best possible trade-off solutions. 
Solving such multi-objective optimization problems directly with a variational quantum algorithm has not been possible until now.

While most investigations focus on qubit quantum computation, qubits can lead to significant encoding overhead for integer optimization problems \cite{yarkoni2021,ChancellorDomainWall2019, Sawaya2023encodingtradeoffs}. 
In the NISQ-era of quantum computation, such overheads may be prohibitive to obtaining useful results. 
It has been suggested some of these limitations can be handled effectively by utilizing multi-level systems called qudits. 
There are indications that qudit systems may provide algorithmic benefits for circuit simplification~\cite{wang_qudits_2020,quditsSimplify2008,Gao2023roleofentanglement}, error correction~\cite{qudit_errorCorr2013,qudit_errorCorr2015} or algorithm speed-up~\cite{gedik_qudit_speedup2015}. 
Qudit-based approaches have been used for several applications like max-cut, graph-coloring,  constraints handling or  machine learning  problems~\cite{bottrillQutrit2023,bravyi2022,deller_quantum_2023, wachDRULQudit2022,rocaJeratquditML2023,qudit_KDE_2022,mikelEVqudits2023, vargas_TSP_qudits2021,bottarelliConstraints2024}.
Generally, qudits represent a very promising route to efficient quantum information processing in enlarged Hilbert spaces. 
~\cite{ringbauerIonQudits2022,ringbauer_entanglement2023, chi_quditprocessor2022,qutrit_experiement2021,Gao2023roleofentanglement}.    
Therefore, we formulate and evaluate the proposed quantum multi-objective optimization algorithm with qudit variables.

The structure of the paper is as follows. After a brief introduction to multi-objective optimization, we repeat the basics of qudit-based formulation of quantum algorithms. Then, we describe the proposed quantum multi-objective (QMOO) algorithm and introduce several benchmark functions for evaluating its performance.  
We present empirical results from algorithm simulations. Finally, we discuss the results, present our conclusions from this work, and give an outlook for future directions of investigations. 

\section{Related Work}

A recent proposal for solving multi-objective optimization problems with NISQ devices~\cite{chiew2023scalarization} utilizes a scalarizing approach. 
The objectives are combined into several single objectives by weighted sums. 
Each scalarized objective is then solved using a regular QAOA to obtain one Pareto-optimal solution. 
Combining the solutions of several different scalarizing functions allows the reconstruction of convex parts of the Pareto-front.

Another recent approach uses the non-dominated sorting genetic algorithm II (\texttt{NSGA-II}) to obtain the solutions to a constraint single-objective optimization problem~\cite{diezvalle2023multiobj-constraints}.
The quantum circuit generates one quantum state from which the classical cost function to be minimized is calculated.
The fraction of constraint-fulfilling solutions in the quantum state is also calculated as a second objective. 
The multi-objective \texttt{NSGA-II} approach is only employed to handle the constraints. 
Finding a single optimized minimal cost solution is targeted. 
Another approach~\cite{mog_vqe2020} utilizes the \texttt{NSGA-II} algorithm to find the best trade-offs between circuit complexity (measured in the number of \texttt{CNOT} gates) and achievable minimal energy solution for each structure. 
In this setup, a full variational single-objective optimization of the quantum circuit is done for a given circuit structure. 
In contrast, in our approach, all different objectives enter the quantum circuit, and the whole set of Pareto-optimal solutions is extracted from the resulting one quantum state at once.

\section{Multi-Objective Optimization}\label{sec:multi-objective-optimization}
A  multi-objective optimization problem (MOP) involves minimizing multiple cost functions $C_k(\mathbf{x})$ simultaneously, i.e., $\vec{C}=(C_1, \ldots, C_K)$ with  $C_k:\mathbf{x}\in\mathcal{X}\rightarrow y_k\in \mathbb{R}$, $k\in\{1,\dots,K\}$. 
For every $\vec{y}^{(1)}$ and $\vec{y}^{(2)}\in \mathbb{R}^K$, we say $\vec{y }^{(1)}$ weakly dominates $\vec{y}^{(2)}$ (denoted by $\vec{y}^{(1)} \preceq \vec{y}^{(2)}$) iff.\ $y^{(1)}_k \leq y^{(2)}_k$, $\forall k\in\{1,\dots,K\}$. 
The Pareto order $\prec$ on $\mathbb{R}^K$ is defined: $\mathbf{y}^{(1)} \prec \mathbf{y}^{(2)}$ iff.\ $\vec{y}^{(1)} \preceq \vec{y}^{(2)}$ and $\vec{y}^{(1)} \neq \vec{y}^{(2)}$.
A point $\mathbf{x} \in \mathcal{X}$ is (Pareto-)efficient iff.\ $\nexists \mathbf{x}'\in\mathcal{X}: (\vec{C}(\mathbf{x}') \prec \vec{C}(\mathbf{x}))$. 
The set of all efficient points in $\mathcal{X}$ is called the \emph{efficient set} or \emph{Pareto set} denoted by $\mathcal{PS}$.
The image of the efficient set under $\vec{C}$ is called the \emph{Pareto front}.

Arguably, the most important performance metric in multi-objective optimization is the well-known hypervolume (HV) indicator~\cite{ZitzlerTLFF03,coello2007evolutionary,BeumeFLPV09,Beume09,GuerreiroFP21}. 
HV of a set $Y\subseteq \R^K$ is defined as the $K$-dimensional Lebesgue measure $\lambda$ of the subset dominated by $Y$ and bounded above by a reference point $\vec{r}\in\R^K$, i.e., $\operatorname{HV}(Y,\vec{r})=\lambda(\{\vec{y}\in\R^K \colon \vec{y} \prec \vec{r} \wedge \exists\vec{p}\in Y, \vec{p}\prec\vec{y}\})$. 
The HV indicator is Pareto-compliant~\cite{ZitzlerTLFF03,ZitzlerBT06,Falcon-CardonaM21,Falcon-CardonaE22}: for two subsets $A$ and $B \subseteq \R^K$, $A\prec B \implies \HV(A,\vec{r}) > \HV(B, \vec{r})$, given the same reference point $\vec{r}$.
Due to this property, the HV indicator is often employed as the performance indicator of multi-objective optimization algorithms.
For continuous problem domains, i.e., $\mathcal{X}\subseteq \R^N$, maximizing the HV indicator leads to an asymptotic distribution of points on the Pareto front~\cite{auger2009hypervolume}, which depends on the curvature thereof. To our knowledge, no theoretical results have been reported for discrete problems. 
In this work, we deal with unconstrained combinatorial integer optimization problems where $\mathcal{X}\subseteq \{0,\dots,d-1\}^N$.

\section{Qudit Representation}\label{sec:qudits}
In this section, we briefly explain qudit quantum computing and motivate its usefulness in integer optimization settings. In integer combinatorial optimization, the goal is to find a specific search variable vector that yields points on the Pareto front of the optimization problem. The search variable $\mathbf{x}\in\mathcal{X}$ is a vector of $N$ finite-ranged integers: $\mathbf{x}=(x_1,\dots,x_N)^\top$, where each variable takes a value $x_n\in\{0,1,\dots,d-1\}$.

In the quantum formulation, these integer variables are represented by a set of $N$ qudits, i.e.,\  $d$-level quantum systems.  
The computational basis is chosen such that each possible search vector corresponds to one $N$ qudit quantum state,
\begin{align}
    \mathbf{x}=(x_1,\dots,x_N)^T \Leftrightarrow \ket{\mathbf{x}}=\ket{x_1,\dots,x_N}
\end{align}
The total dimension of the Hilbert space is $\text{dim}\, \mathcal{H} = d^N$, and any state vector $\ket{\psi} \in \mathcal{H}$ can be written as
\begin{align}
\ket{\psi} = \sum_{x_1=0}^{d-1} \ldots \sum_{x_N=0}^{d-1}  \alpha_{x_1 \ldots x_N} \ket{\mathbf{x}} \, ,
\end{align}
where $\alpha_{x_1 \ldots x_N}$ is the complex amplitude with $\sum_{\mathbf{x}}|\alpha_{\mathbf{x}}|^2=1$ and the states $\ket{\mathbf{x}} = \ket{x_1, \ldots, x_N }$ form an orthonormal basis, i.e.,  $\braket{\mathbf{x}|\mathbf{x}'}=\delta_{\mathbf{x},\mathbf{x}'}=\delta_{x_1,x_1'}\dots\delta_{x_N,x_N'}$.

In general, one needs $d^2-1$ Hermitian basis operators to generate all unitary operations of the $ SU (d)$ group, which represent all allowed actions on a single qudit in the quantum circuit~\cite{wang_qudits_2020}.
However,  it was shown~\cite{kasper_universal_2022, giorda_universal_2003} that it is sufficient to consider only the three operators $L_x$, $L_z$, and  $L_z^2$ as generators. 
Here, $L_x$ and $L_z$ are the  $x$- and $z$-angular momentum operators, which act on the single qudit state as follows,
\begin{align}
\label{eq:angularmometum}
L_x\ket{x} & = \tfrac12\big(\gamma_{d,x+1}\ket{x+1}+\gamma_{d,x-1}\ket{x-1}\big) \\
L_z\ket{x} & = \tfrac{2x-d+1}{2}\ket{x}
\end{align}
where $x\in\{0,\dots,d-1\}$ and  $\gamma_{d,x}=\sqrt{(d-x-1)(x+1)}$. 
The operator $L_z^2$ is called squeezing or one-axis twisting operator in the context of cold-atom systems. 
These three generators are sufficient to implement arbitrary unitary operations by (possibly many) repeated finite rotations. 
This is because the iterated commutators of these three operators generate all $d^2-1$ Hermitian basis operators, which are then necessary to generate all unitary operations of the group $ SU (d)$. 

\section{Quantum Multi-Objective Optimization}\label{sec:quantum-multiple-objective-algorithm}
In this section, we explain the construction of our algorithm to solve multi-objective combinatorial optimization problems natively on quantum hardware.

Each objective $C_k(\textbf{x})$ is represented by a cost Hamiltonian $H_k$, which is diagonal in the computational basis $\ket{\textbf{x}}$. 
The  eigenvalue of the basis state is given by the cost function value of that state,
\begin{align}
\label{eq:cost_ham}
    H_k \ket{\mathbf{x}}&=C_k(\mathbf{x})\ket{\mathbf{x}}\quad (k=1,\dots,K).
\end{align}

\begin{figure*}
    \centering
    \includegraphics[width=0.7\linewidth]{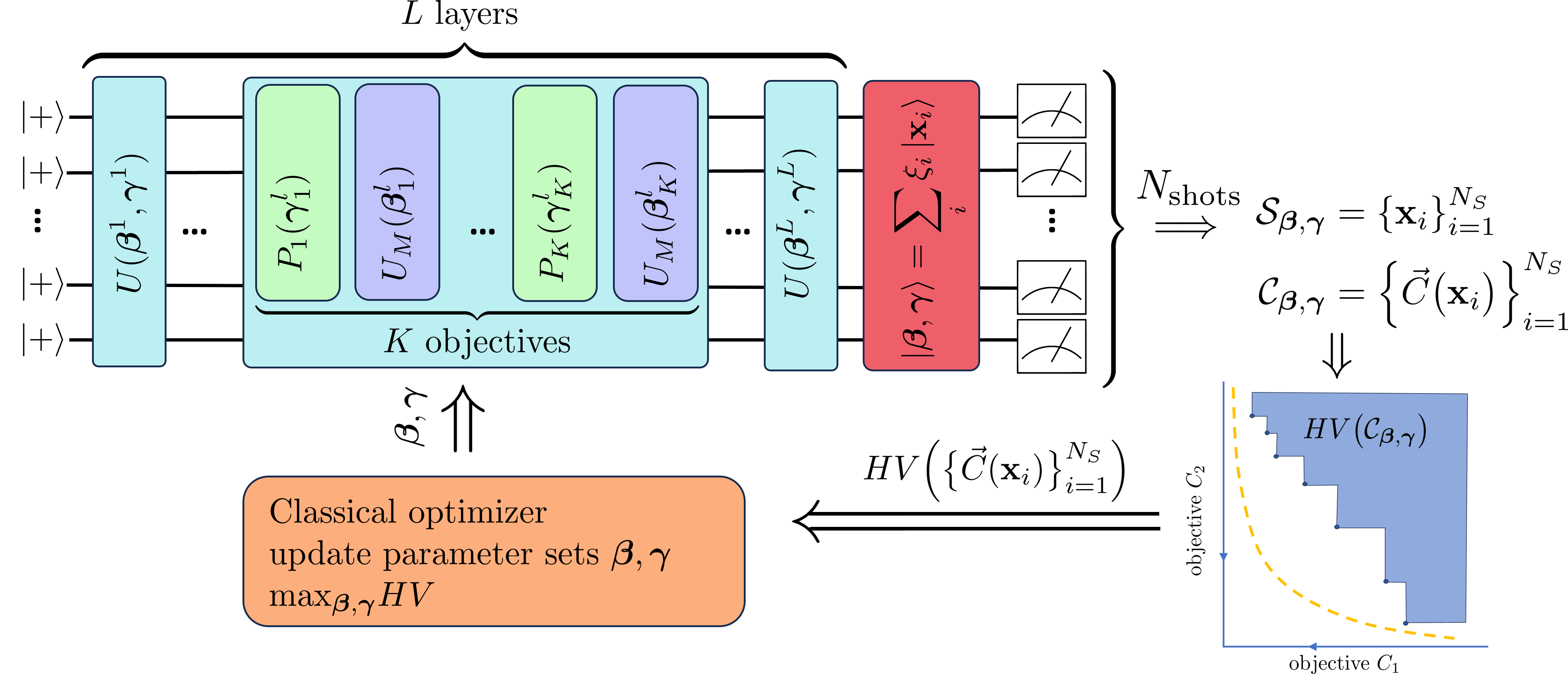} 
    \caption{Schematic overview of the QMOO algorithm. The algorithm is initialized in an even superposition state. The circuit consists of $L$ layers (blue), each of which consists of phase- (green) and mixing-operators (purple) for each of the $K$ objectives. 
    The QMOO state (red) is measured $N_\text{shots}$ times, and a population containing $N_S$ solutions is generated from the output. 
    We calculate the hypervolume from this population, which is used as a metric for the classical optimizer tuning the parameter sets $\mathbf{\beta}, \mathbf{\gamma}.$ 
    The updated parameters are used in the next iteration's QMOO circuit, and the quantum-classical loop is completed.
    }
    \label{fig:algorithm-schematic}
\end{figure*}

We employ a variational quantum circuit to solve many-objective optimization problems. A schematic can be seen in Fig.~\ref{fig:algorithm-schematic}.
Analogous to the traditional QAOA, the circuit is built up from $L$ layers, where the functional form of the unitary operation in each layer is the same across all layers, but each layer has independently adjustable variational parameters $\boldsymbol{\beta}^l$ and $\boldsymbol{\gamma}^l$ with $l=1,..,L$,
\begin{align}
    \label{eq:circuit_state}
    \ket{\psi(\boldsymbol{\beta}, \boldsymbol{\gamma})}  &=  U(\boldsymbol{\beta}^L,\boldsymbol{\gamma}^L) \cdots U(\boldsymbol{\beta}^1,\boldsymbol{\gamma}^1)\ket{\psi_0}\;.
\end{align}
$\ket{\psi_0}$ denotes a predefined fixed initial state. We choose the uniform superposition of all possible states, 
\begin{equation}
    \ket{\psi_0} = \frac{1}{\sqrt{d^N}}\sum_{\mathbf{x}\in \{0, \cdots, d-1\}^N}\ket{\mathbf{x}},    
\end{equation}
which is an eigenvector to the generalized Pauli $X$-operator~\cite{wang_qudits_2020}.
Alternative initial states are also possible, such as one single basis state or an eigenstate to the $x$ angular momentum operator $L_x$. 
The latter is typically done for qubit circuits.
However, the specific choice of the initial state should not change the algorithm's performance qualitatively: the different initial states are related by unitary transformations, which can be generated implicitly by the variation quantum circuit. 
Therefore, the initial states merely alter the variational optimization problem slightly without changing its characteristics. 
We experimentally validated this argument with various initial states (not shown in this paper).  

Each layer $l$ of the circuit comprises $K$ blocks, where $K$ is the number of classical objectives. 
Each block consists of one mixing unitary $U_M$ and one phase unitary $P_k$,  
\begin{align}
\label{eq:layerunitary}
    U(\boldsymbol{\beta}^l, \boldsymbol{\gamma}^l) &=  U_M(\boldsymbol{\beta}^l_K) P_{K}(\boldsymbol{\gamma}^l_K) \cdots  U_M(\boldsymbol{\beta}^l_1) P_{1}(\boldsymbol{\gamma}^l_1) 
\end{align}
This is in contrast to the regular QAOA, where only one phase and one mixing operator are included, i.e.\ $K=1$.

While, in principle, the mixing unitaries of each block in Eq.~\eqref{eq:layerunitary} could have a different structure, we choose them to be the same but allow for different variational parameters of each mixer.
We focus on mixing unitaries which are generated by the $x$-angular momentum $L_x$ and the local squeezing operator, $L_z^2$, which is sufficient to generate all unitaries as described in Sec.~\ref{sec:qudits} above. 
This leads to the mixing unitary of block $k$
\begin{align}
    U_M(\boldsymbol{\beta}^l_k)  &= e^{-i \beta^l_{1,k} \sum_{j=1}^N  L_{x;j}  -i \beta^l_{2,k} \sum_{j=1}^N  L_{z;j}^2}  \,,
\end{align}
where the sum in the exponent extends over all qudits $j=\{1,\dots,\dots,N\}$. 
Note that for qubits ($d=2$), the local squeezing operator reduces to the identity, $L_z^2=\tfrac14\mathds{1}$, and consequently, we drop that term for $d=2$. 

The phase operators are generated by the $K$ cost Hamiltonians representing the classical objectives, i.e.,\
\begin{align}
    P_{k}(\boldsymbol{\gamma}^l_k)  &= e^{-i \gamma^l_k H_k}\,.
\end{align}
where $k=\{1,\dots,K\}$.

For a given set of variational parameters $(\boldsymbol{\beta},\boldsymbol{\gamma})$ the quantum circuit generates a state as depicted in Eq.~\eqref{eq:circuit_state}.  
Generally, this state is a superposition of all computational basis states,  i.e., all classical solutions, with varying amplitudes.
A set of solutions is generated by running the quantum circuit $N_{\text{shots}}$ times and measuring one output state each time. 
From this set,  the $N_S$ most frequent basis states  are selected 
\begin{align}
    \ket{\psi(\boldsymbol{\beta}, \boldsymbol{\gamma})} 
    \overset{N_{\text{shots}}}{\Longrightarrow }
    \mathcal{S}_{\boldsymbol{\beta},\boldsymbol{\gamma}} = \{\mathbf{x}_i\}_{i=1}^{N_S} \,.
\end{align}
In case of infinite number of shots, $N_{\text{shots}}\to\infty$, this is analogous to performing a full state tomography (see, e.g., ~\cite{barnett_quantum_2009,kurmapuTomography2023,parisTomography2004}) and selecting the $N_{S}$ states with largest spectral weights $|\xi_{\mathbf{x}}|^2$.
But as we show, the algorithm works well with a finite and small number of shots, much less than the dimension of the Hilbert space.

For each of those extracted classical solutions, all classical cost objectives are calculated, which generates the set of objective vectors: 
\begin{align}
\mathcal{C}_{\boldsymbol{\beta}, \boldsymbol{\gamma}} & = \{\{C_k(\mathbf{x}_i)\}_{k=1}^K\}_{i=1}^{N_S}\\
               & = \{\vec{C}(\mathbf{x}_i)\}_{i=1}^{N_S}\,.
\end{align}
From this set, we extract the non-dominated solutions, which approximate the Pareto front. 
The quality of the current approximate solution set is then evaluated by calculating the hypervolume indicator with respect to a predefined reference point $\vec{r}$ 
\begin{align}
    \label{eq:hypervolume}
    \HV\left(\{\vec{C}(\mathbf{x}_i)\}_{i=1}^{N_S}; \vec{r}\right).
\end{align}
This serves as the single classical objective function, which is maximized to determine the optimal values of the parameters $(\boldsymbol{\beta},\boldsymbol{\gamma)}$.

The idea behind this choice of the unitaries in each layer is that each phase operator produces relative phases proportional to the energies of the corresponding objective. 
In that way, the information of each objective is encoded in relative phases of states, contributing to the quantum state prepared by the circuit.
Given the individual variational parameters for each objective, the classical optimization procedure can, therefore, in principle, discriminate between states based on the set of objective values and find optimal parameters to increase the weights of Pareto-efficient solutions in the quantum state.

The ideal quantum state of the proposed multi-objective approach is given by a superposition where  Pareto optimal solutions are the predominant contribution.
That is, the ideal state  represents the whole set of classical solutions belonging to the Pareto set $\mathcal{PS}$ as one superposition, i.e.\
\begin{align}
\label{eq:qmoo_final_state}
    \ket{\psi(\boldsymbol{\beta}_\text{ide}, \boldsymbol{\gamma}_\text{opt})}_{\mathrm{ideal}} = \sum_{\textbf{x}\in\mathcal{PS} } \xi_\textbf{x} \ket{\textbf{x}}+\ket{\delta\psi},
\end{align}
with $\sum_{\mathbf{x}\in\mathcal{PS}} |\xi_\mathbf{x}|^2\approx1$ and $\braket{\delta\psi|\delta\psi}\ll 1$.
This represents a major difference to current variational algorithms like QAOA, which target reducing the quantum state to one optimal solution, i.e., one basis state. 

The total number of variational parameters of the proposed algorithm is $N_\text{params}=3LK$, where $L$ is the number of layers and $K$ is the number of objectives.
For the case of qubits ($d=2$) without a local squeezing operator, the number of parameters reduces to $N_\text{params}=2LK$.
In both cases, the number of parameters is increased by a factor of $K$ compared to the standard QAOA.

\section{Benchmark Functions and Experimental Setup}\label{sec:benchmarks_setup}
\subsection{Benchmark Functions}\label{sec:benchmarks}
The performance of the proposed multi-objective quantum optimization algorithm is evaluated on several benchmark functions. 
We combine linear and quadratic cost functions into two-, three-, and five-objective problems where all objectives are subject to minimization.
To define a useful multi-objective benchmark, the objectives should be in a trade-off relation, where improving on one objective typically leads to performance degradation in other objectives. 
The benchmark functions are available as Python functions from a \texttt{github} repository~\cite{qmoo_benchmarks}.

The first class of benchmark problems is defined by a linear two-objective function of the form 
\begin{align}
    \label{eq:linbench}
    \vec{C}_\mathrm{I}(\mathbf{x})&=
    \begin{pmatrix}
    \mathbf{c}_1\cdot \mathbf{x}\\
    \mathbf{c}_2\cdot \mathbf{x}
    \end{pmatrix}\,.
\end{align}
Here, $\mathbf{x}\in\{0,\dots,d-1\}^N$ and the cost coefficients $\mathbf{c}_1$ are sampled from $[-1,1]^N$ uniformly at random, i.e., $\mathbf{c}_1\sim \mathcal{U}([-1, 1]^N)$.
The trade-off between the objectives is generated by choosing the second cost coefficients to have substantial anti-correlations with the first coefficients, i.e.,\ 
$\mathbf{c}_2\sim-\frac12\mathbf{c}_1+\frac12 \mathcal{U}([-1,1]^N)$.

We also composed two-, three-, and five-objective benchmark problem classes using randomized quadratic cost functions of the form  
\begin{align}
\label{eq:Ising}
    C(\mathbf{x})&=
    \mathbf{x}^\top  \mathbf{J}\mathbf{x} +\mathbf{h}^\top \mathbf{x}\,,
\end{align}
where  $\mathbf{J}=\mathbf{J}^\top\in \mathbb{R}^{N\times N}$   are symmetric randomized coupling matrices and $\mathbf{h}\in\mathbb{R}^N$ are randomized local fields. 
To generate trade-offs between the objectives, we are guided by physical spin models that realize such cost Hamiltonians. 

For one two-objective problem class ($K=2$), we chose one objective to favor ferromagnetic (FM) alignment of the variables and the other to favor anti-ferromagnetic (AFM) alignment.  
Consequently, the symmetric coupling matrix elements are chosen from uniform distributions, i.e., $J_{ij}=J_{ji}\sim \mathcal{U}(a,b)$ where for the FM case all matrix elements are negative, i.e.,\ $a<b<0$, and for the AFM case all are positive, i.e.,\ $b>a>0$. For the cost function to truly represent spin models, the local fields are chosen as $\mathbf{h}=\mathbf{g}-d\mathbf{1}^\top\cdot\mathbf{J}$ where $\mathbf{g}$ is a random vector with $g_i\sim \mathcal{U}(-1,1)$, $d$ the local dimension of the qudits and $\mathbf{1}$ the $N$-dimensional vector of ones.
We refer to this as problem class II and denote the objective function vector by $\vec{C}_\text{II}$.

The third benchmark problem class also has two quadratic objectives. The first is implemented with AFM couplings described above, and the second is designed to minimize the distance of the randomly scaled search variables to a random point $\mathbf{x}_0\sim \mathcal{U}([0,d-1])^N$. 
Consequently the cost function matrix is diagonal with randomized positive entries, i.e.,\ $J_{ij}=\delta_{ij} J_r$ where $J_r \sim\mathcal{U}(a,b)$ with $0<a<b$, and the local fields are determined by the scaled random point $\mathbf{h}=-2\mathbf{x}_0^\top\cdot\mathbf{J}$. 
We denote this as problem class III with objective vectors $\vec{C}_\text{III}$.

The fourth problem class denoted as $\vec{C}_\text{IV}$ are three-objective problems ($K=3$). The first and second objectives are AFM and FM-coupled spin models as in problem $\text{II}$, respectively, and the third objective is given by the quadratic distance to a given random vector $\textbf{x}_0$ as described for the second objective of problem class III.   

The last class of benchmark problem has five objectives and is denoted by $\vec{C}_\text{V}$. 
The first two objectives are chosen to be the AFM and FM spin chains with nearest neighbor coupling, i.e.,\ $J_{ij}=J_{ji}\sim \delta_{j,i+1}\mathcal{U}(a,b)$ with  $b>a>0$ and $a<b<0$ for the AFM and the FM case, respectively. The third objective is the distance to a given random vector  $\mathbf{x}_0\sim \mathcal{U}([0,d-1]^N)$, i.e.,\ $J_{ij}=\delta_{ij}$ and $\mathbf{h}=-2\mathbf{x}_0$. 
The fourth (fifth) objective is also realized as a nearest neighbor spin-chain, but where the first half of the variables, $x_q$ with $q\leq N/2$, have FM (AFM) couplings while the latter half, $x_q$ with $q>N/2$, have opposite, i.e.\ AFM (FM), couplings. 

\begin{figure}
  \includegraphics[width=0.6\linewidth]{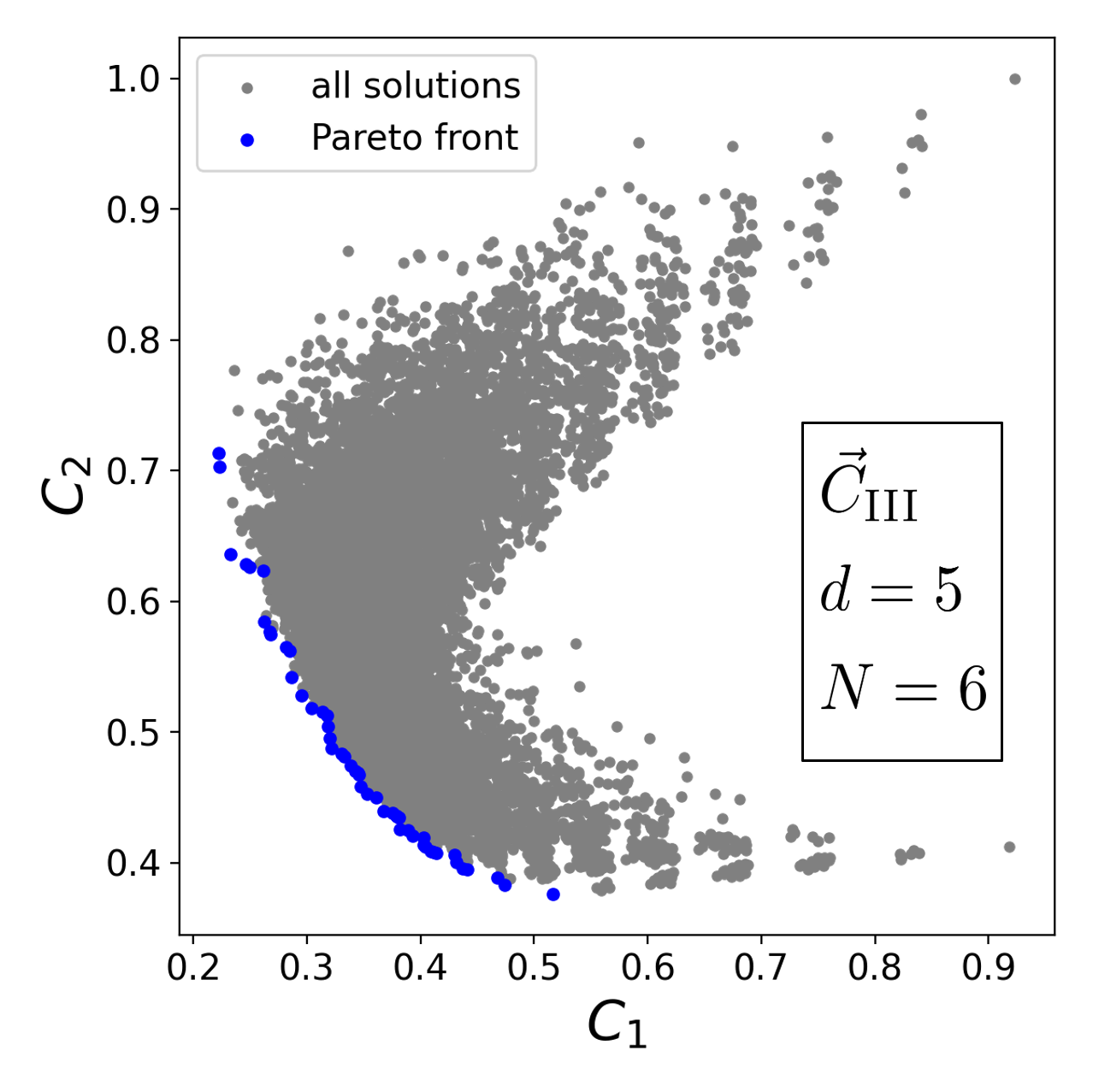} 
    \caption{Objective space of one example problem instance with $N=6$ ququints ($d=5$) of the benchmark problem type  $\vec{C}_\text{III}$. 
    The gray dots indicate all possible solutions, while the blue dots indicate the Pareto-optimal front.}
    \label{fig:single_problem}
\end{figure}

Figure \ref{fig:single_problem} shows the objective space of one specific instance of the bi-objective problem class $\vec{C}_\text{III}$. The gray dots represent all possible solutions while the blue dots indicate the Pareto-optimal front.  
Note that we normalized each objective cost function such that the values are between $C(\textbf{x})\in [0,1]$ for all functions, numbers of qudits $N$, and dimensionality of the qudits $d$.  
This allows us to always take the reference vector for the hypervolume calculation as the all-ones vector $\vec{r}=(1,\dots,1)^\top$.
More example instances of benchmark problems are found in Figure~\ref{fig:problems_schematic} in Appendix~\ref{sec:app_bench}.

\subsection{Experimental Setup}
\label{sec:setup}

The proposed quantum multi-objective optimization algorithm is evaluated on several instances of the previously introduced benchmark functions. For each specific randomly generated instance of a benchmark problem, the algorithm is run 50 times with different random initializations of the parameter values of the quantum circuits and the classical optimizers.
For each run, all parameters are initialized randomly in the interval $[-\pi,\pi]$.
In each iteration of the classical optimization algorithm,  $N_{\text{shots}}$ measurement shots are performed on the quantum state prepared by the quantum circuit.
From the measurement results the $N_S=20$ most frequent  classical solutions are selected. 
We vary $N_{\text{shots}}$ between 128 and $\infty$ to investigate the dependency of the algorithm on measurement shot noise, which is a necessary step toward bringing the algorithm to real hardware.  

We tested several classical optimization algorithms to determine the optimized parameters of the quantum circuit. We used COBYLA, L-BFGS-B, and Powell from the \texttt{scipy} library~\cite{virtanen_scipy_2020}, covariance matrix adaptation evolutionary strategies (CMA-ES)~\cite{hansenCMA} taken from~\cite{CMAESRepo}, and in-house implementations of particle swarm optimization (PSO)~\cite{pso1995}, and differential evolution (DE)~\cite{de_2997}. 
The overall performance of all tested optimization algorithms was roughly comparable to each other over all settings and benchmarks. An exception was the L-BFGS-B algorithm, which yields consistently lower hypervolume values, most likely because it is a quasi-Newton method with a numerical evaluation of the gradients. 
The population-based stochastic optimizers like PSO, CMA-ES, and DE gave slightly better results than Powell and COBYLA in most instances. 
However, given their large resource requirements, which are long algorithm runtimes and a large number of circuit evaluations, we focus on Powell algorithm.

For comparison, we also run several state-of-the-art classical multi-objective evolutionary optimization algorithms (MOEAs) from the \texttt{Platypus}~\cite{platypus} package. Specifically, these are \texttt{NSGA-II}, \texttt{IBEA}, and \texttt{MOEA/D}.
We set the population size of the MOEAs to 20, comparable to $N_S=20$ extracted solutions from the quantum states in QMOO. 
We execute each MOEA with 40 independent runs on each benchmark problem and limit the number of iterations to 200.

\section{Results}
\begin{figure*}[!ht]
  \includegraphics[width=0.24\linewidth]{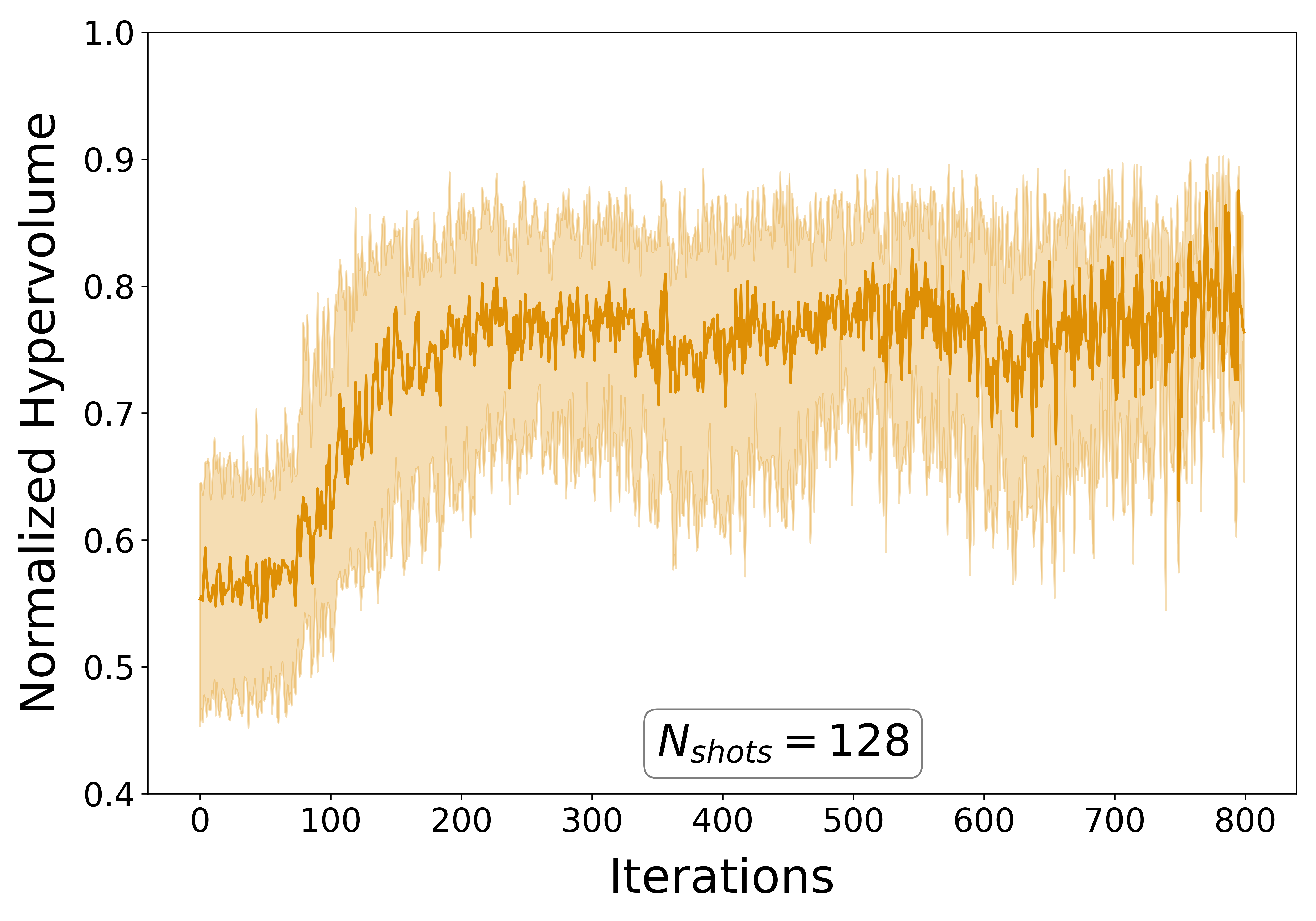} 
  \includegraphics[width=0.24\linewidth]{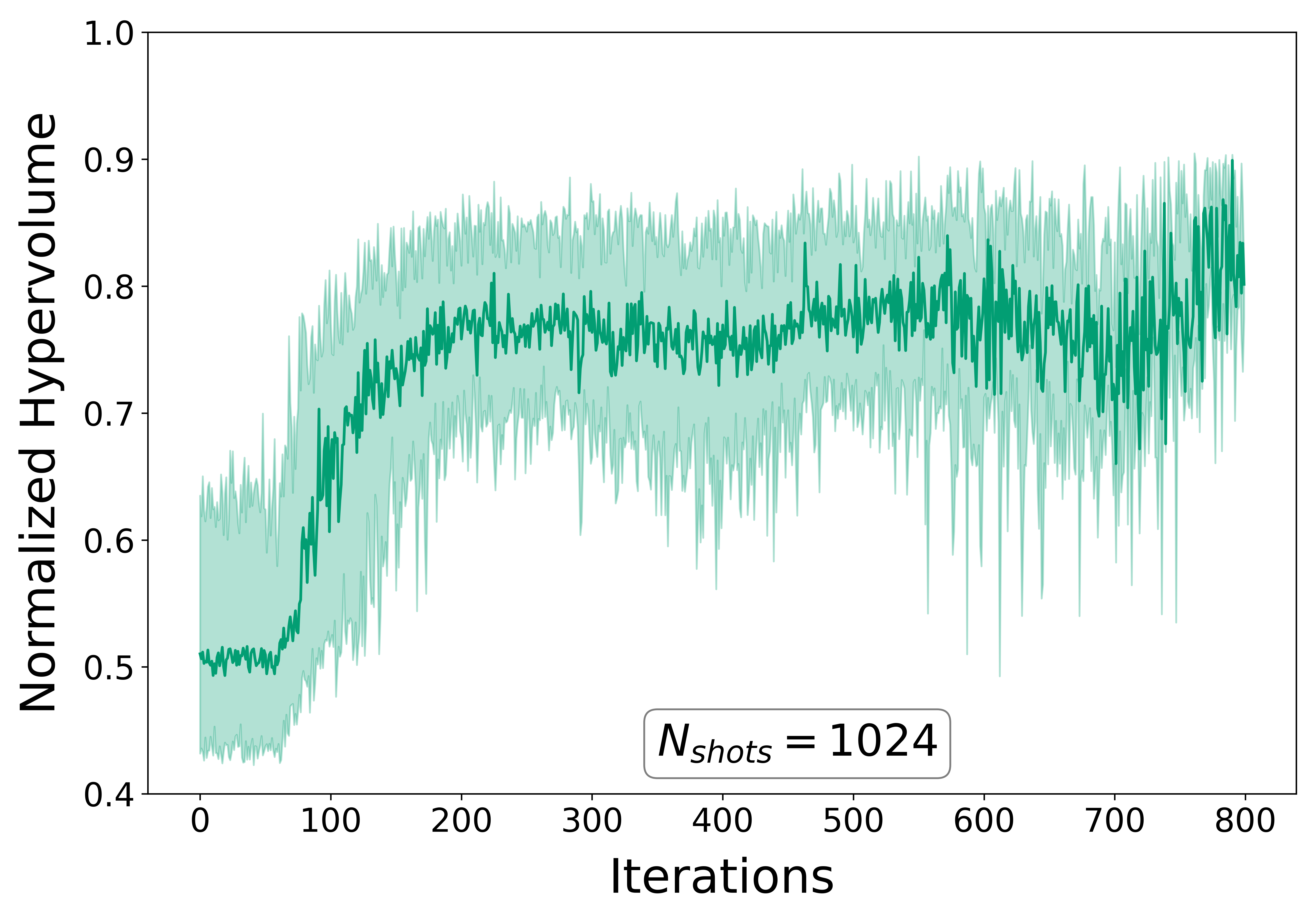} 
  \includegraphics[width=0.24\linewidth]{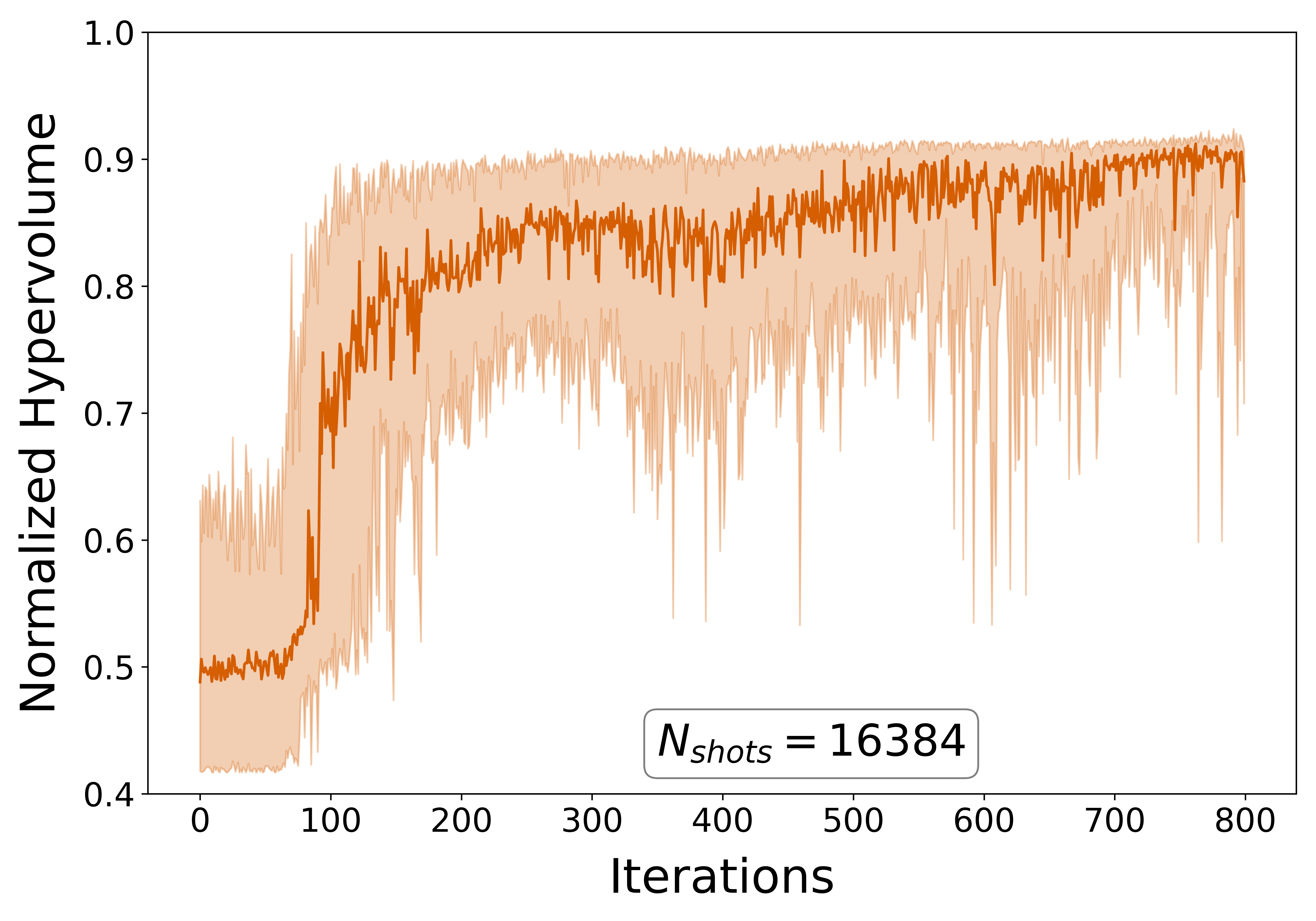} 
  \includegraphics[width=0.24\linewidth]{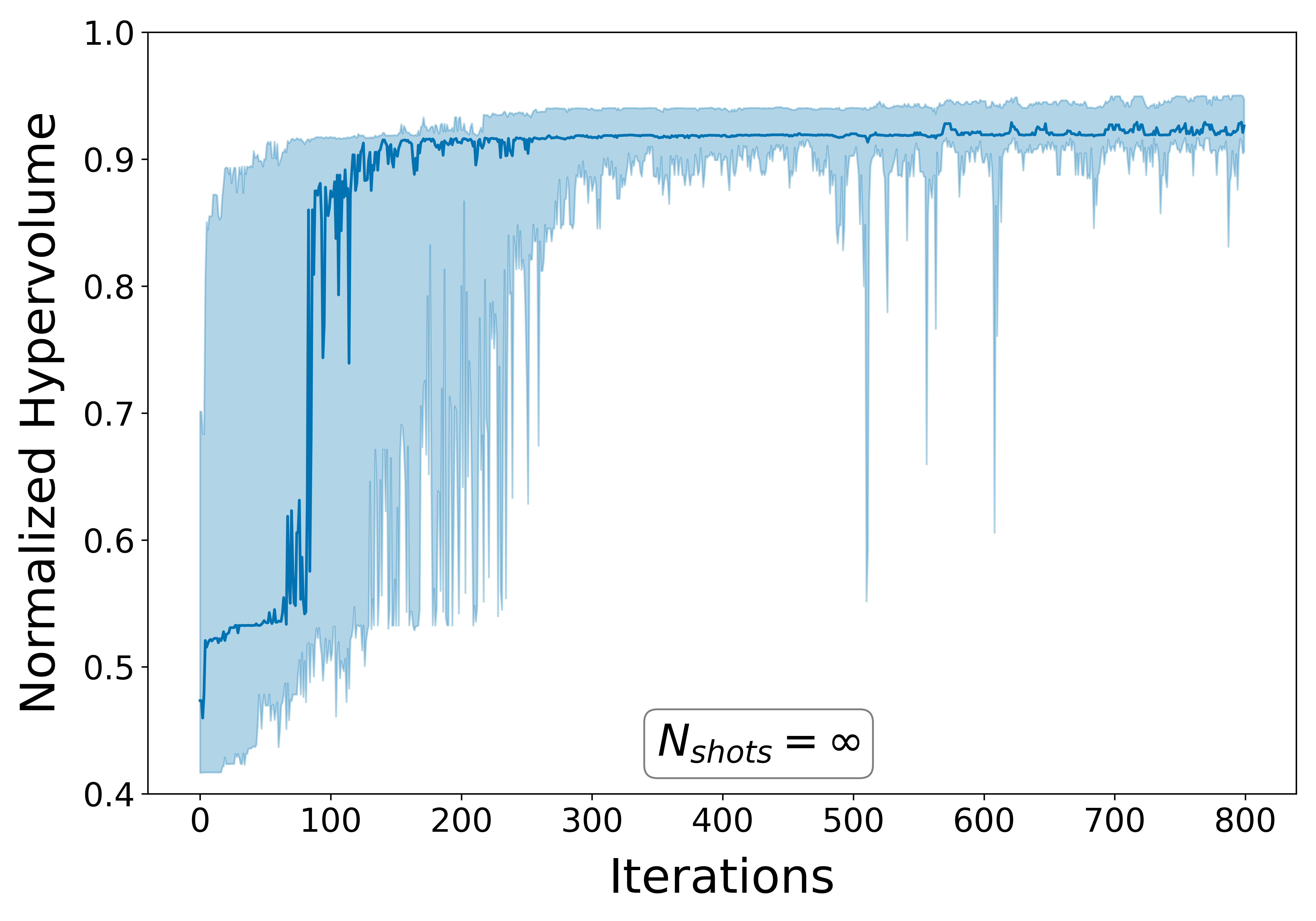} 
    \caption{Normalized hypervolume, i.e., \ hypervolume relative to the maximal achievable value of the Pareto front, as a function of the iterations of the QMOO algorithm for one particular instance of the five objective optimization problem using $N=6$ ququints ($d=5$).  
    The thick line represents the median taken over 50 runs with different random initialization of the QMOO parameters, and the shaded region denotes the $20$th and $80$th quantiles. The classical Powell optimizer is used to optimize the quantum circuit consisting of $L=1$ layer. 
    The results are for increasing the number of measurement shots $N_{\text{shots}}$ during the QMOO iterations (from left to right): $N_{\text{shots}}=128$, $N_{\text{shots}}=1024$, $N_{\text{shots}}=16384$,  and $N_{\text{shots}}=\infty$ (full state tomography). 
    For all the measurement shots, the performance increases with several iterations. 
    The lower number of shots $N_{ahots}$ leads to more noise, but already rather small numbers of shots, such as 128, give satisfactory results.
    }
    \label{fig:problem-overview}
\end{figure*}

Figure~\ref{fig:problem-overview} shows the evolution of the hypervolume as a function of the classical optimization iterations for 50 runs of one specific instance of a five-objective benchmark problem $\vec{C}_\text{V}$.
The algorithm increases the hypervolume over the iterations for all considered numbers of measurement shots $N_{\text{shots}}$. 
Even for as low as $N_{\text{shots}}=128$ shots, it increases, and the algorithm produces increasingly better sets of Pareto efficient solutions.
For lower $N_{\text{shots}}$
the progress of the hypervolume is more noisy. In the context of optimization, this is not necessarily a drawback. More noise during the optimization implies more exploration in the search space, which can be beneficial or detrimental depending on the classical optimization algorithm. 

\begin{figure}[!ht]
    \centering
    \includegraphics[width=0.6\linewidth]{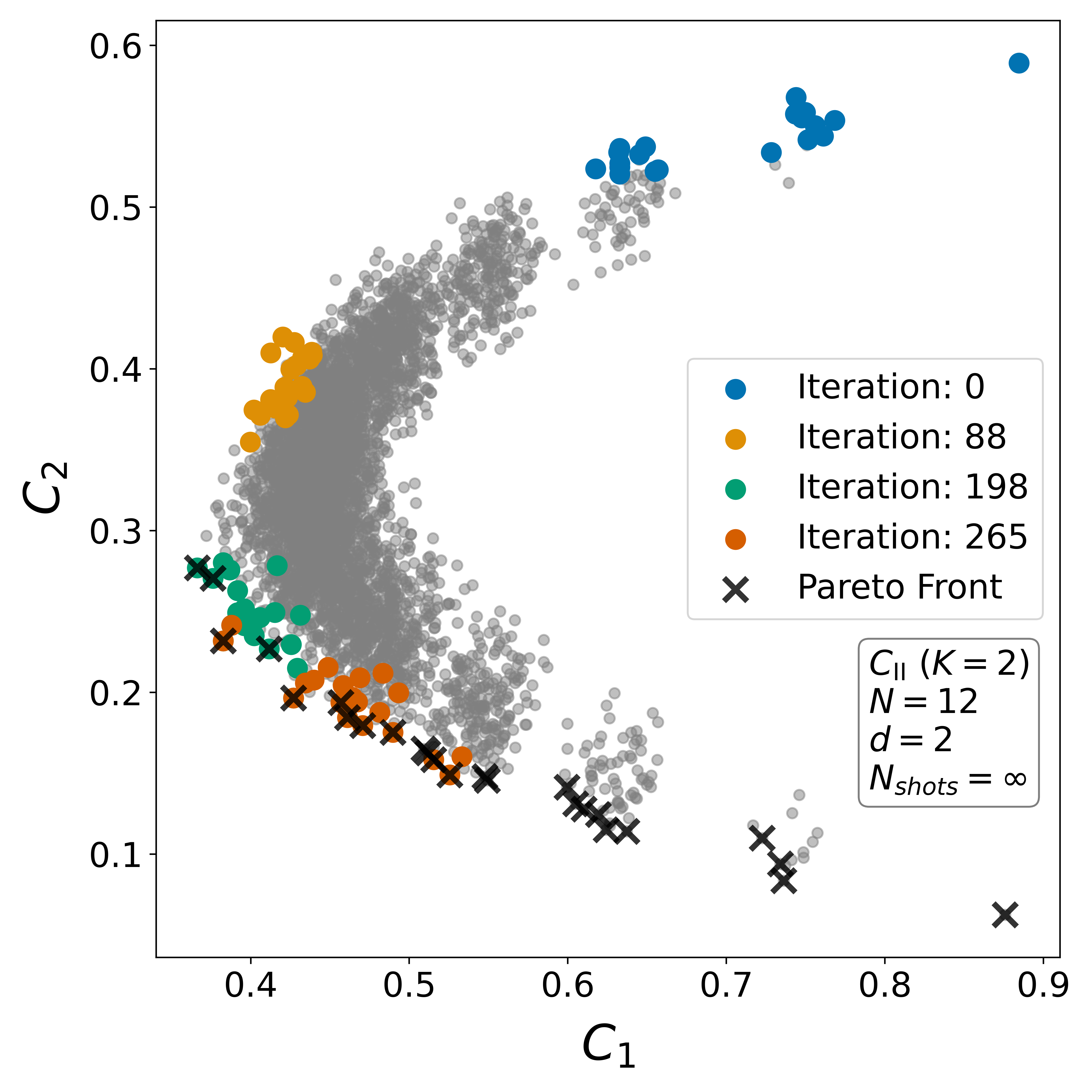} 
    \caption{Progression of the solution set of one run for a $L=1$ layer QMOO circuit of the $K=2$ objective quadratic problem of class $C_\text{II}$ for $N=12$ qubits. 
     Gray dots show all possible solutions, while the colored dots show the solutions extracted from the QMOO at the indicated iterations. The Pareto front is shown with black crosses.
     Over the iteration, the solutions generated by the QMOO approach move toward the Pareto front.}
    \label{fig:objective-space}
\end{figure}

The increase in the hypervolume indicates the algorithm's efficiency in generating increasingly better approximations of Pareto-efficient solution sets.
To explicitly show this, we show an example of progress for one specific optimization run for a two-objective quadratic problem of class $\vec{C}_\text{II}$ in Fig.~\ref{fig:objective-space}. 
The two-objective problem was chosen for visualization purposes. 
The random initialization leads to initial solutions in the top right part of the objective space.
With increasing iterations, better parameters of the QMOO circuit are found, and the solution set represented in the final quantum state progresses toward the Pareto front.

\begin{figure*}
    \includegraphics[width=0.32\linewidth]{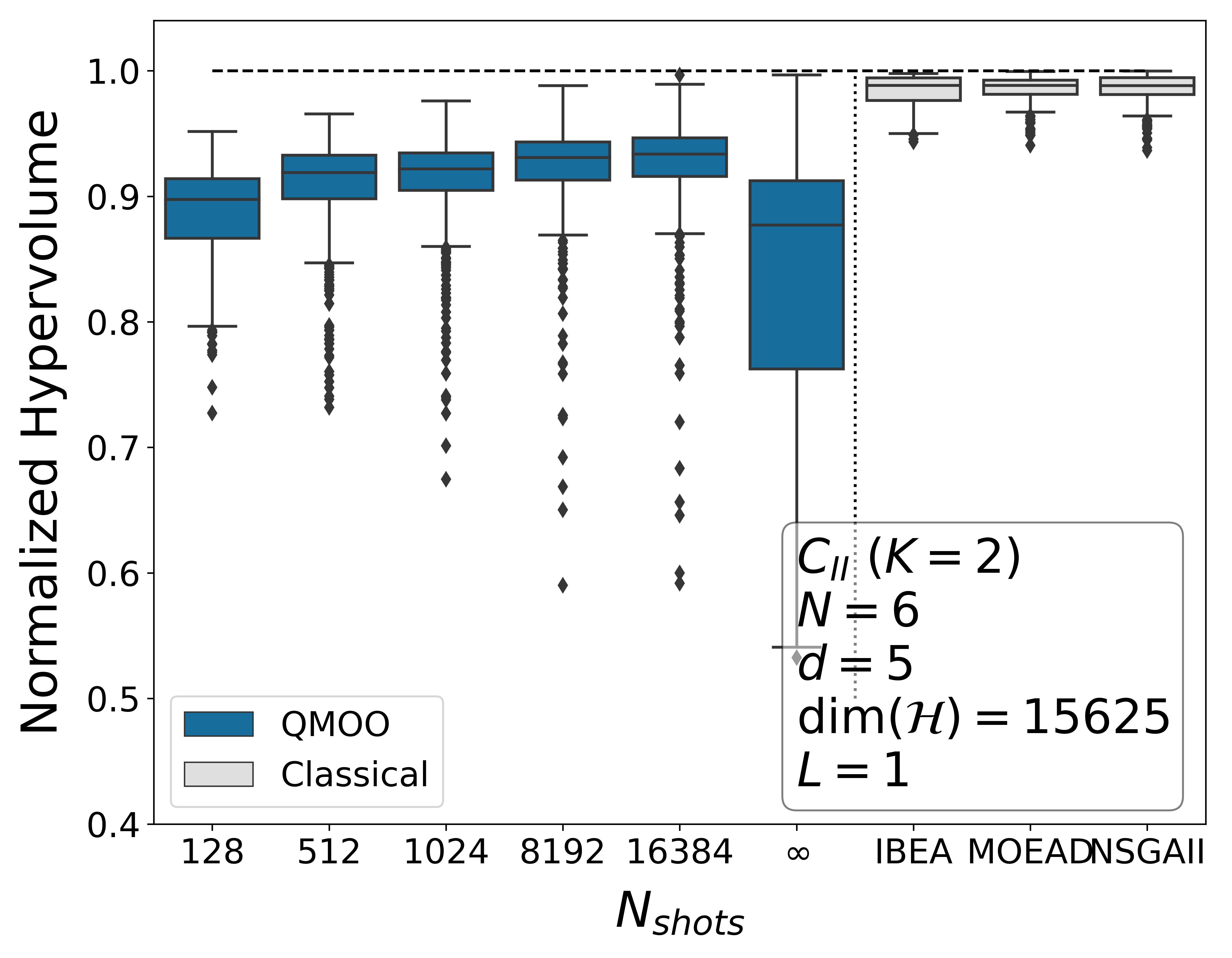} 
    \includegraphics[width=0.32\linewidth]{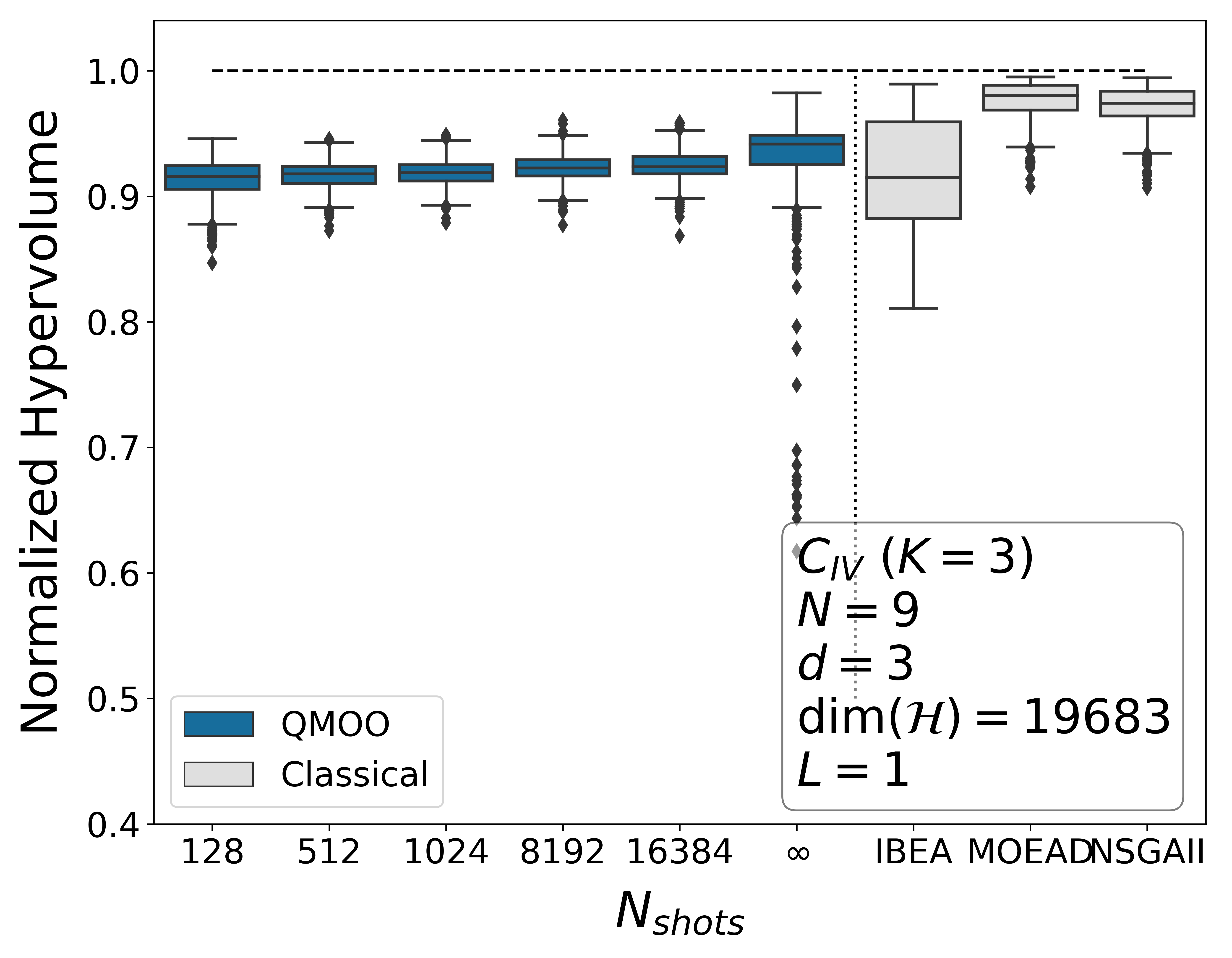} 
    \includegraphics[width=0.32\linewidth]{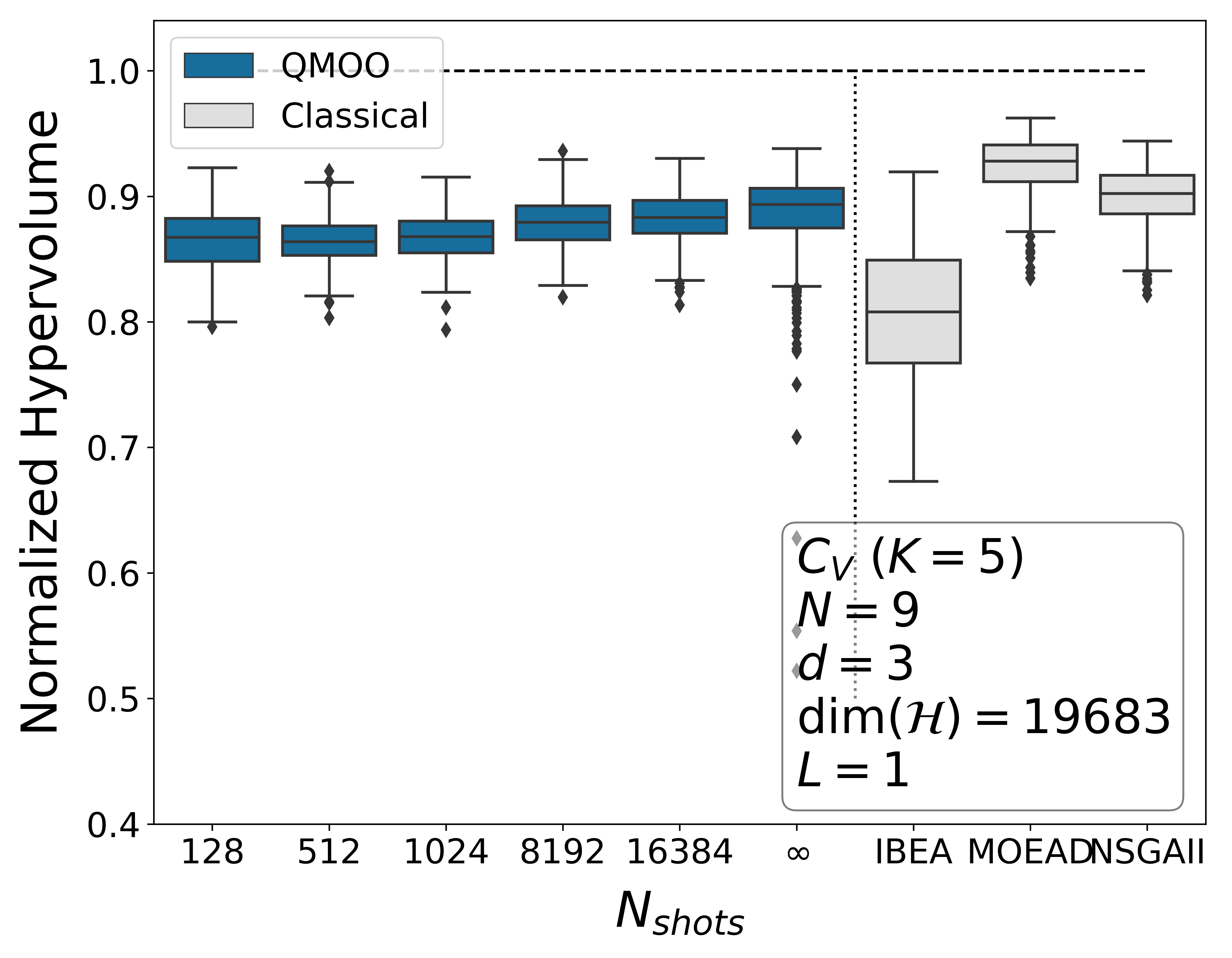} 
    \caption{Normalized hypervolume for $L=1$ layer QMOO circuit as a function of shots for three problem classes. Left: $N=6$ ququints ($d=5$) on bi-objective problem class $\vec{C}_\text{II}$. 
    Middle:  $N=9$ qutrits ($d=3$) for the the three-objective problem class  $\vec{C}_\text{IV}$. 
    Right: $N=9$ qutrits ($d=3$) for the the five-objective problem class  $\vec{C}_\text{V}$.
    The results are from 11 problem instances and 50 randomly initialized runs for each problem instance. 
    In each panel, results from three classical multi-objective algorithms are shown for comparison on the right in grey. 
    The dependence of the QMOO results on the number of measurement shots is rather weak, and good results are already achieved for a low number of shots, much smaller than the Hilbert space dimension, which is of the order of $10^4$ for these problems.  
    The QMOO performance reaches up to the classical optimizers, having a slightly lower median for bi-objective problems (left) but comparable for three and five objectives (middle and right). 
    }
    \label{fig:HV-as-a-function-of-shots} 
\end{figure*}

When targeting to run the algorithm on actual hardware, it is mandatory to explore its behavior when changing the number of measurement shots $N_{\text{shots}}$ performed in each iteration of QMOO to extract the $N_S$ solutions.
The results presented in Fig.~\ref{fig:HV-as-a-function-of-shots} show that the dependency on $N_{\text{shots}}$ is overall rather weak.
The algorithm's performance is almost unchanged when increasing the shots from as low as $128$ to as high as $16384$. 
Even performing exact state tomography ($N_{\text{shots}}=\infty$) gives similar results.
For some problem classes, the performance seems to slightly increase with an increasing number of shots, while all results are within the statistical variation of each other.
However, for problem class $C_\text{II}$ with two objectives, it seems that exact full-state tomography gives more statistical variations and even slightly worse results than a finite number of shots.
This hints at the fact that for this specific problem class, the noise induced by the measurement shots leads to a more explorative search in the classical optimization of the circuit, which seems to be beneficial for the optimization procedure.

In Fig.~\ref{fig:HV-as-a-function-of-shots}, we  compare the performance of three classical multi-objective optimizers. 
For the two-objective problems, the proposed QMOO approach reaches up to but is still mostly inferior in performance to the classical approaches. At the same time, it is comparable to the three and five-objective problem classes.
Also, it should be noted that the variation between the classical optimizers is also rather large for the three and five objective problems. 
This is a sign of the well-known fact that the performance of optimization algorithm can be very problem dependent and there is not one globally best algorithm for all problems \cite{wolpertNoFreeLunchLearning96,wolpertNoFreeLunchOpt97}.

\begin{figure}[!ht]
    \centering
    \includegraphics[width=\linewidth]{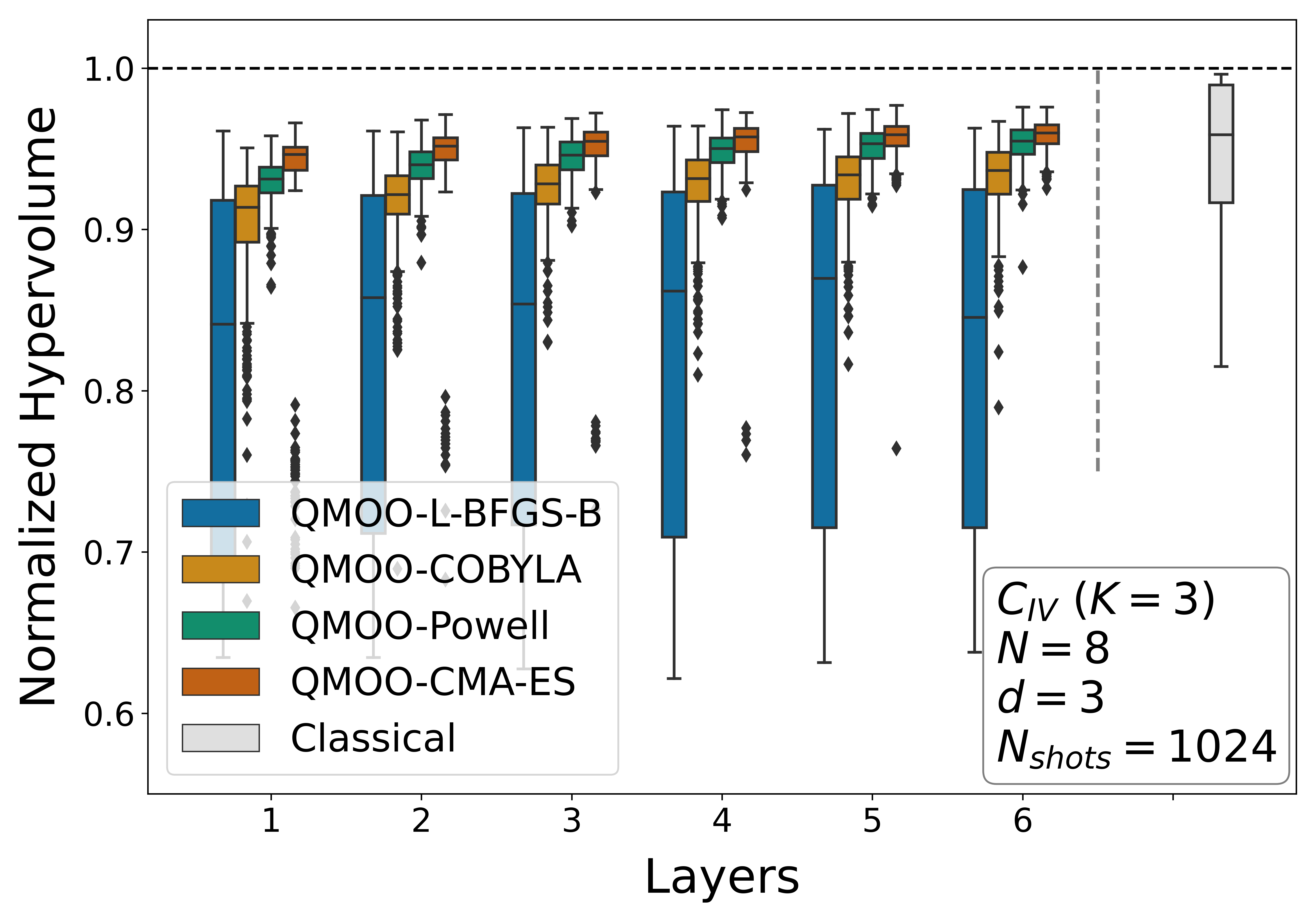} 
    \caption{Normalized hypervolume as a function of  QMOO circuit layers for different classical optimizers for the problem class $C_\text{IV}$ with $K=3$ objectives for $N=8$ qutrits ($d=3$) using $N_\text{shots}=1024$ shots. 
    The Hilbert space dimension for these problems is dim$(\mathcal{H})=6561$. 
    The statistics are collected from 11 problem instances and 50 runs for each instance.  
    The results from the purely classical optimization are shown on the right.
    The classical optimization algorithms can be clearly ranked from L-BFGS-B over COBYLA and Powel to CMA-ES (worst to best). 
    The hypervolume increases with more layers for all cases, and the values are comparable to the classical approaches for this example.}
    \label{fig:HV-layers-optmizer}
\end{figure}

The effect of the choice of the classical optimizer and the influence of number of layers $L$ in the quantum circuit were also investigated. Representative result are shown in Fig~\ref{fig:HV-layers-optmizer} for  the problem class $\vec{C}_\text{IV}$. We observed qualitatively similar results for all other problems and parameter choices.

The performance of all tested optimization algorithms can be clearly ranked.   
The CMA-ES gives the best results, while  Powell and COBYLA are  following closely. 
The L-BFGS-B approach gives significantly worse results with very large statistical variations. 
This is most likely a consequence of using numerically evaluated gradients, which are rather unstable for noisy cost functions.
While achieving the best performance, the CMA-ES has the major drawback that it needs an order of magnitude more circuit evaluations since it operates with a population of solutions candidates (in our case 10), which need to be evaluated in each optimization iteration. 
Therefore, for our main results we used the Powell optimizer as it represents a very good trade-off between achievable results quality and required computational effort. 

Figure~\ref{fig:HV-layers-optmizer} also shows that the QMOO results are comparable to  the classical optimization algorithms for these problem instances.  
The performance increases as a function of the layers in the quantum circuit,  as expected. 
However, this increase is only minor. 
This was the motivation to focus only on single-layer ($L=1$) circuits in this proof-of-concept study.

\begin{figure*}
     \includegraphics[width=0.32\linewidth]{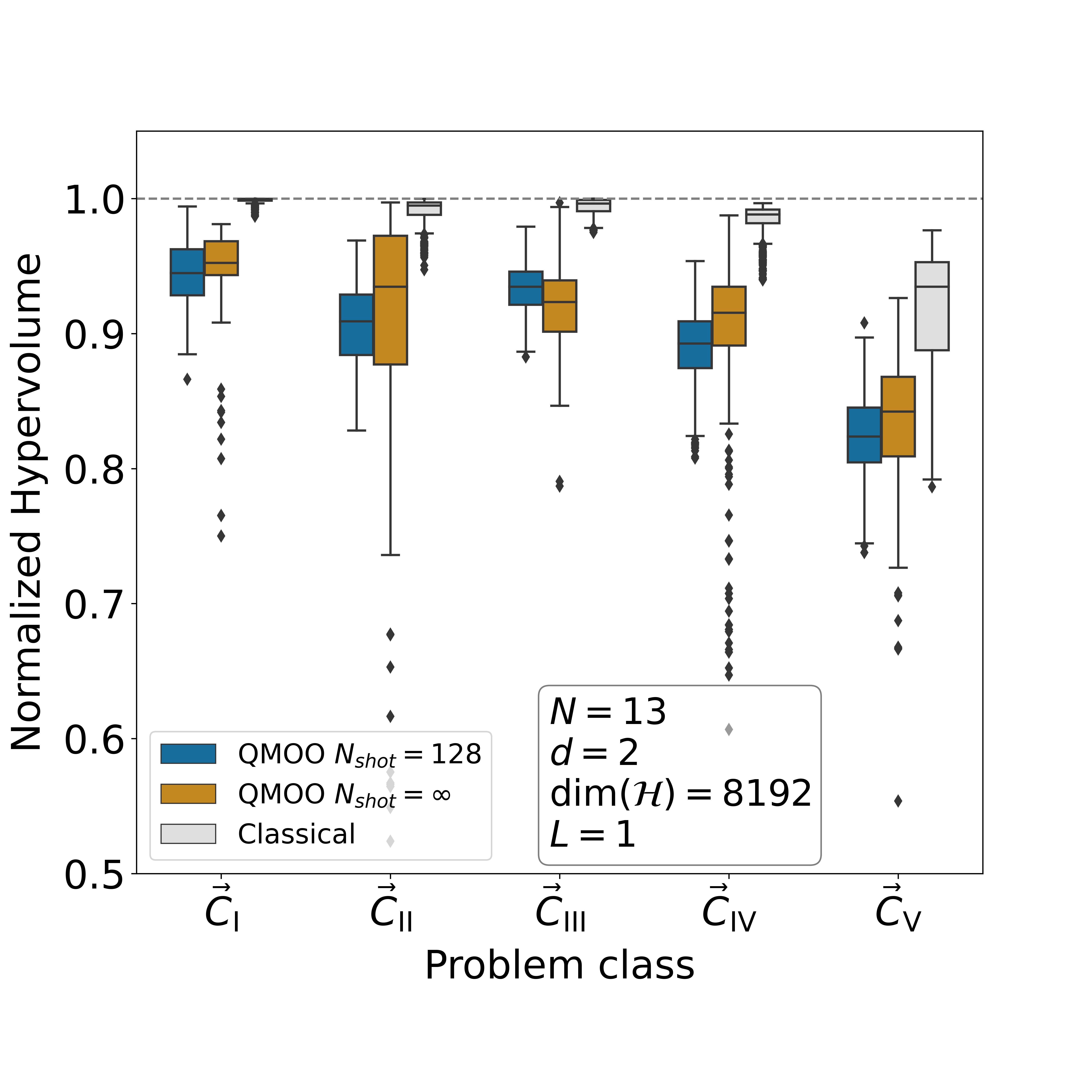} 
     \includegraphics[width=0.32\linewidth]{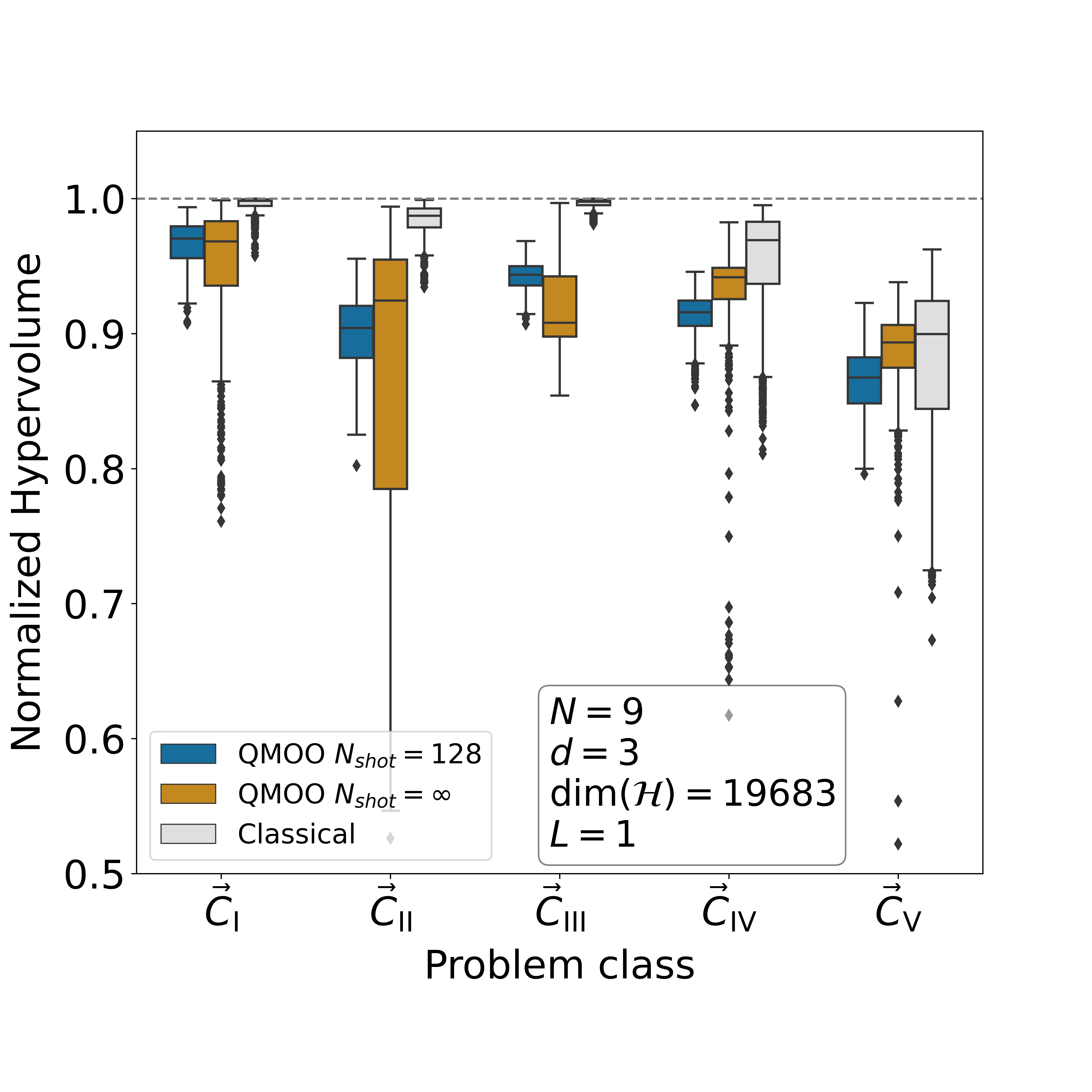}
     \includegraphics[width=0.32\linewidth]{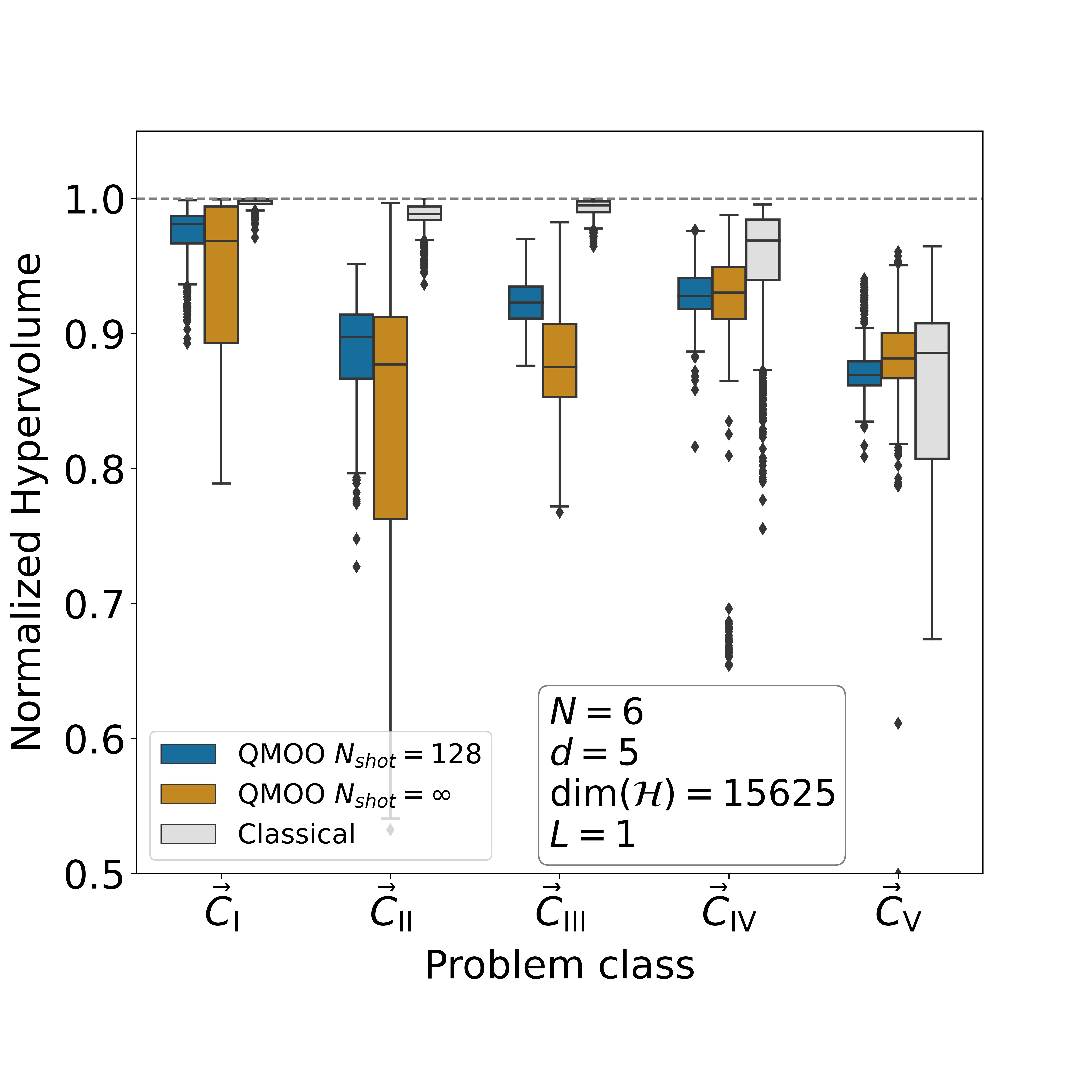} 
    \caption{Normalized hypervolume of the QMOO approach using $N_{\text{shots}}=128$ measurement shots (blue)  and exact state tomography, i.e.\ $N_{\text{shots}}=\infty$ (orange) on all benchmark problem classes for different setups compared to the results from purely classical multi-objective solvers (grey). Left: $N=13$ qubits ($d=2$).  Middle: $N=9$ qutrits ($d=3$). Right:  $N=6$ ququints ($d=5$).
    The quantum circuits only consisted of $L=1$ layer and were optimized with the Powell optimizer. 
    The results are for 11 different problem instances for each problem class and 50 randomized runs for each specific problem instance. 
    The QMOO approach gives satisfactory results already with low number of shots and even reaches comparable performance to classical optimizers for the three and five objective problem classes, $\vec{C}_\text{IV}$  and $\vec{C}_\text{V}$ respectively.  
    }
    \label{fig:true-pareto-solutions}
\end{figure*}

We summarize the overall results of the proposed QMOO algorithm with $L=1$ layer in Fig.~\ref{fig:true-pareto-solutions} 
We show the performance for all five benchmark problem classes for a setups with $N=13$ qubits ($d=2$), $N=9$ qutrits ($d=3$), and $N=6$ ququints ($d=5$). 
We include results from QMOO using only 128 measurement shots and when using full-state tomography ($N_{\text{shots}}=\infty$).
We also compare the statistics of the achieved hypervolume values to the ones obtained with purely classical algorithms. 
The first observation is that the conclusions drawn from the previous analysis regarding the dependency on the number of shots hold for all studied problem setups.
The performance with $N_{\text{shots}}=128$ is similar to that obtained using exact state tomography. 
For problem class $C_\text{III}$, it even seems to be slightly larger with a lower number of shots, indicating enhanced performance due to more explorative search as a result of increased shot noise. 
Compared to the classical approaches, the QMOO approach performs slightly lower for problems with $K=2$ objectives but is comparable for the three and five-objective problem classes. 
It should be noted that classical multi-objective algorithms are the result of considerable research efforts and have improved substantially over time by implementing best practices and adaptation strategies. 
In contrast, no such tuning or improvement has yet been made to the QMOO approach.
Therefore, we may expect the QMOO performance to improve in the future by hyper-parameter tuning~\cite{sanderAlgo2017,stuetzleIRACE2016} or warm-starting~\cite{Egger2021warmstartingquantum,akshayParameterConQAOA2021, sackQAOA_init2023,Jain2022graphneuralnetwork,bravyiRecursive2020,blekosQAOA_Review,scrivaVaraitionalShots2024}, which is known have a potentially large impact.

\section{Discussion and Conclusion}

The proposed QMOO algorithm utilizes quantum superpositions as a resource to represent a whole set of classical solutions in one quantum state. 
We showed that this allows us to solve multi-objective optimization problems by optimizing a parametrized quantum circuit to maximize the hypervolume indicator for the extracted classical solutions. 
Unlike the standard single objective QAOA, the algorithm does not target the quantum state generated by the variational circuit to reduce to a single computational basis state (i.e.,\ one classical solution). 
Rather, the goal of the QMOO is to obtain a quantum state consisting of a superposition of non-dominated solutions, ideally belonging to the  Pareto front of the multi-objective optimization problem. 

We showed the effectiveness of the QMOO approach by evaluating its performance over a variety of benchmark problem classes. 
These classes are built from randomized linear and quadratic cost functions, which can be formulated with an arbitrary number of qudit variables.
Throughout the variational optimization iterations, the hypervolume indicator increases, indicating a quality improvement of the approximation to the Pareto-front. 
The proposed QMOO approach was able to find solutions of similar quality to the purely classical methods for many cases.  
But overall, the performance is not yet fully competitive with state-of-the-art classical multi-objective algorithms.

The results suggest that the  QMOO algorithm is inherently resilient to shot noise and does not require exponential scaling of the number of measurement shots with Hilbert space dimension. 
We evaluated this for Hilbert space dimension up to dim$(\mathcal{H})\approx 2\cdot10^4$ and could not detect a systematic degradation in perfromance with lower number of shots.
As few as 128 measurement shots per iteration already gave consistently good results.  
This is reassuring in the light of running this algorithm on  quantum hardware in the near future. 

We are aware that our analysis merely constitutes a first step, and many issues need to be investigated in the future. 
In the context of qudit quantum circuits, a potential performance increase can be realized by including more generators of the $ SU (d)$ group into the mixing operators~\cite{wachDRULQudit2022}.
Similarly, allowing each qudit to have an individual mixing parameter to increase the flexibility of the quantum circuit was shown to be beneficial in some applications for QAOA~\cite{herrmanMA-QAOA_2022, saleem_localMixingQAOA2021, shi_localMixerQAOA_2022}. 
Including more than one layer, as we focused on in this work, is also expected to increase performance as indicated by our result shown in Fig.~\ref{fig:HV-layers-optmizer}. 
However, these measures come at the cost of greatly increasing the number of variational parameters and leading to a more complicated classical optimization problem.
Finally, random initialization and direct optimization of all parameters of the circuit as we currently do it is not very efficient and much more favorable schemes exist~\cite{Egger2021warmstartingquantum,akshayParameterConQAOA2021, sackQAOA_init2023,Jain2022graphneuralnetwork,bravyiRecursive2020,blekosQAOA_Review,scrivaVaraitionalShots2024}.

A very important topic to consider is trainability and the so-called barren plateau phenomenon~\cite{laroccaBarrenPlateauReview2024}. 
On the one hand, the structure of each layer in the proposed QMOO algorithm is more complex than the standard QAOA, as there is one mixing and one phase operator for each objective. 
As circuit complexity is one route to barren plateaus,  the QMOO circuit could be suspected to be prone to trainability issues for deep circuits with many layers. 
On the other hand, the QMOO approach does not need to calculate an expectation value with the output quantum state of the circuit. 
Rather, individual basis states are sampled from the quantum state and processed, which is also the reason for the resilience of the QMOO algorithm to a low number of measurement shots. 
Thus, one prominent source of trainability issues and barren plateaus is eliminated. 
The investigation of these issues is left for future research. 

In summary, the proposed QMOO algorithm is a very promising novel approach to solving multi-objective optimization problems natively on quantum hardware. 
The approach requires few measurement shots and is, as such, suitable for running on actual quantum hardware. 
We are confident that this approach will widen the application domain for quantum-optimization approaches and hopefully allow for successful solutions to real-world problems of practical relevance. 
We consider this work a first step toward a truly quantum algorithm for solving multi-objective optimization problems, which utilizes quantum resources for computing Pareto-optimal sets.

\section{Acknowledgements}
We acknowledge fruitful discussions with T.~Bäck, X.~Bonet-Monroig, A.~Bottarelli, V.~Dunjko, P.~Hauke and B.~Sendhoff.

LE \ and SS \ acknowledge funding by the European Union under the Horizon Europe Programme. 
Views and opinions expressed are, however, those of the author(s) only and do not necessarily reflect those of the European Commission. 
Neither the European Union nor the granting authority can be held responsible for them. 
Grant Agreement 101080086 — NeQST.

\appendix

\section{Visualization of exemplary benchmark problems}
\label{sec:app_bench}

\begin{figure*}
  \includegraphics[width=0.94\linewidth]{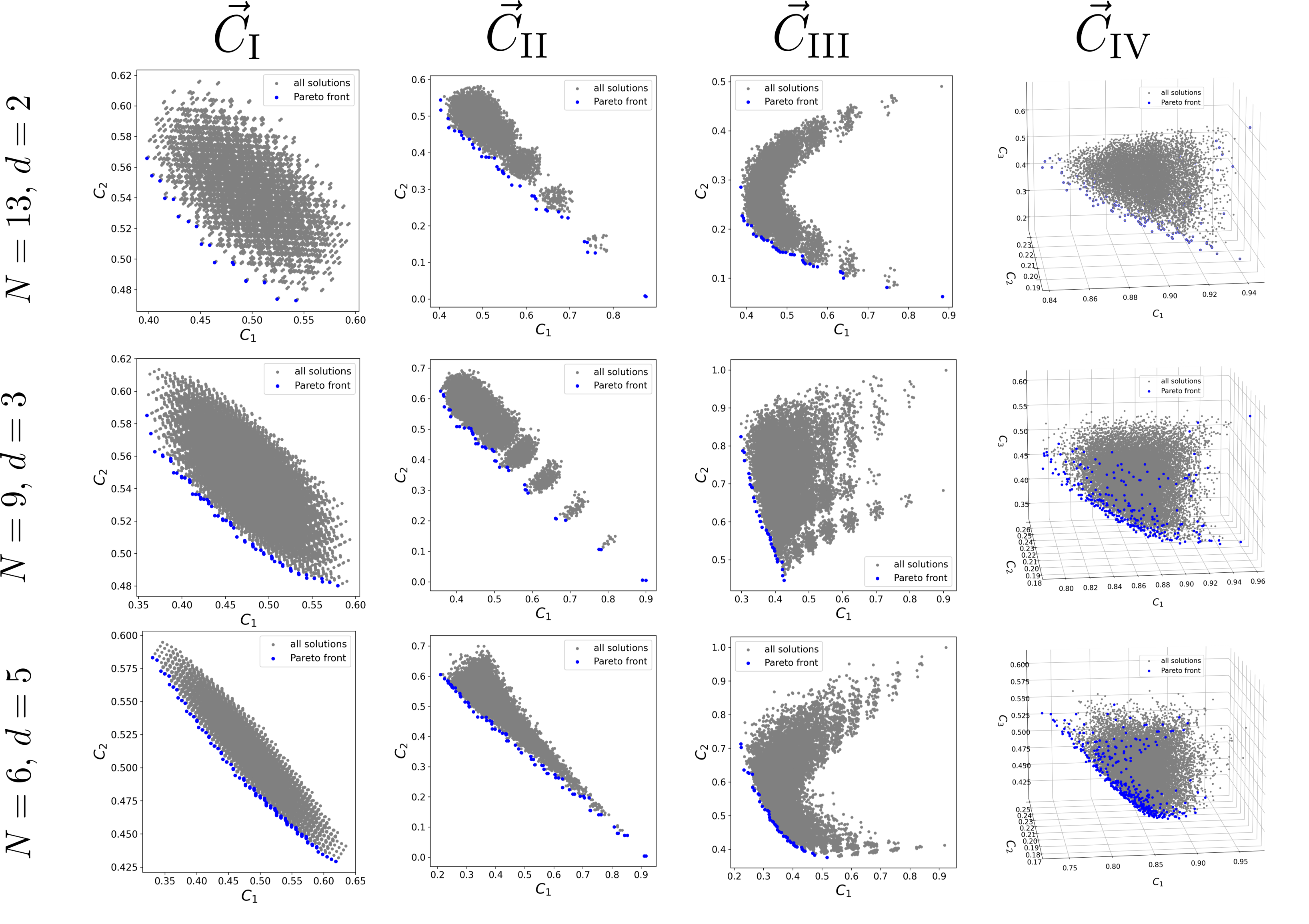} 
    \caption{Objective space of example instances of several benchmark problems.  
    The problems are, from left to right: linear $K=2$ objective $C_\text{I}$, two quadratic $K=2$ problems $C_\text{II}$ and $C_\text{III}$ and a quadratic $K=3$ objective problem $C_\text{IV}$. 
    The top row is for $N=13$ qubits ($d=2$), the middle row is $N=9$ qutrits ($d=3$), and the lowest row is $N=6$ ququints ($d=5$).  The gray dots indicate all possible solutions, while the blue dots indicate the true Pareto-optimal front.}
    \label{fig:problems_schematic}
\end{figure*}

Figure~\ref{fig:problems_schematic} shows schematic examples of the linear and quadratic objective functions for two- and three-objective benchmarks for qubits ($d=2$), qutrits ($d=3$) and ququints ($d=5$).  The gray dots indicate all possible solutions while the blue dots denote Pareto-optimal solutions.


\begin{thebibliography}{74}%
\makeatletter
\providecommand \@ifxundefined [1]{%
 \@ifx{#1\undefined}
}%
\providecommand \@ifnum [1]{%
 \ifnum #1\expandafter \@firstoftwo
 \else \expandafter \@secondoftwo
 \fi
}%
\providecommand \@ifx [1]{%
 \ifx #1\expandafter \@firstoftwo
 \else \expandafter \@secondoftwo
 \fi
}%
\providecommand \natexlab [1]{#1}%
\providecommand \enquote  [1]{``#1''}%
\providecommand \bibnamefont  [1]{#1}%
\providecommand \bibfnamefont [1]{#1}%
\providecommand \citenamefont [1]{#1}%
\providecommand \href@noop [0]{\@secondoftwo}%
\providecommand \href [0]{\begingroup \@sanitize@url \@href}%
\providecommand \@href[1]{\@@startlink{#1}\@@href}%
\providecommand \@@href[1]{\endgroup#1\@@endlink}%
\providecommand \@sanitize@url [0]{\catcode `\\12\catcode `\$12\catcode
  `\&12\catcode `\#12\catcode `\^12\catcode `\_12\catcode `\%12\relax}%
\providecommand \@@startlink[1]{}%
\providecommand \@@endlink[0]{}%
\providecommand \url  [0]{\begingroup\@sanitize@url \@url }%
\providecommand \@url [1]{\endgroup\@href {#1}{\urlprefix }}%
\providecommand \urlprefix  [0]{URL }%
\providecommand \Eprint [0]{\href }%
\providecommand \doibase [0]{https://doi.org/}%
\providecommand \selectlanguage [0]{\@gobble}%
\providecommand \bibinfo  [0]{\@secondoftwo}%
\providecommand \bibfield  [0]{\@secondoftwo}%
\providecommand \translation [1]{[#1]}%
\providecommand \BibitemOpen [0]{}%
\providecommand \bibitemStop [0]{}%
\providecommand \bibitemNoStop [0]{.\EOS\space}%
\providecommand \EOS [0]{\spacefactor3000\relax}%
\providecommand \BibitemShut  [1]{\csname bibitem#1\endcsname}%
\let\auto@bib@innerbib\@empty
\bibitem [{\citenamefont {{Google Quantum AI}}(2023)}]{acharya2022suppressing}%
  \BibitemOpen
  \bibfield  {author} {\bibinfo {author} {\bibnamefont {{Google Quantum AI}}},\
  }\href@noop {} {\bibfield  {journal} {\bibinfo  {journal} {Nature}\ }\textbf
  {\bibinfo {volume} {614}},\ \bibinfo {pages} {676–681} (\bibinfo {year}
  {2023})}\BibitemShut {NoStop}%
\bibitem [{\citenamefont {Kim}\ \emph {et~al.}(2023)\citenamefont {Kim},
  \citenamefont {Eddins}, \citenamefont {Anand}, \citenamefont {Wei},
  \citenamefont {Berg}, \citenamefont {Rosenblatt}, \citenamefont {Nayfeh},
  \citenamefont {Wu}, \citenamefont {Zaletel}, \citenamefont {Temme},\ and\
  \citenamefont {Kandala}}]{Youngseok2023evidence}%
  \BibitemOpen
  \bibfield  {author} {\bibinfo {author} {\bibfnamefont {Y.}~\bibnamefont
  {Kim}}, \bibinfo {author} {\bibfnamefont {A.}~\bibnamefont {Eddins}},
  \bibinfo {author} {\bibfnamefont {S.}~\bibnamefont {Anand}}, \bibinfo
  {author} {\bibfnamefont {K.}~\bibnamefont {Wei}}, \bibinfo {author}
  {\bibfnamefont {E.}~\bibnamefont {Berg}}, \bibinfo {author} {\bibfnamefont
  {S.}~\bibnamefont {Rosenblatt}}, \bibinfo {author} {\bibfnamefont
  {H.}~\bibnamefont {Nayfeh}}, \bibinfo {author} {\bibfnamefont
  {Y.}~\bibnamefont {Wu}}, \bibinfo {author} {\bibfnamefont {M.}~\bibnamefont
  {Zaletel}}, \bibinfo {author} {\bibfnamefont {K.}~\bibnamefont {Temme}},\
  and\ \bibinfo {author} {\bibfnamefont {A.}~\bibnamefont {Kandala}},\ }\href
  {https://doi.org/10.1038/s41586-023-06096-3} {\bibfield  {journal} {\bibinfo
  {journal} {Nature}\ }\textbf {\bibinfo {volume} {618}},\ \bibinfo {pages}
  {500} (\bibinfo {year} {2023})}\BibitemShut {NoStop}%
\bibitem [{\citenamefont {Bluvstein}\ \emph {et~al.}(2023)\citenamefont
  {Bluvstein}, \citenamefont {Evered}, \citenamefont {Geim}, \citenamefont
  {Li}, \citenamefont {Zhou}, \citenamefont {Manovitz}, \citenamefont {Ebadi},
  \citenamefont {Cain}, \citenamefont {Kalinowski}, \citenamefont {Hangleiter},
  \citenamefont {Ataides}, \citenamefont {Maskara}, \citenamefont {Cong},
  \citenamefont {Gao}, \citenamefont {Rodriguez}, \citenamefont {Karolyshyn},
  \citenamefont {Semeghini}, \citenamefont {Gullans}, \citenamefont {Greiner},
  \citenamefont {Vuletić},\ and\ \citenamefont {Lukin}}]{queraLogical2023}%
  \BibitemOpen
  \bibfield  {author} {\bibinfo {author} {\bibfnamefont {D.}~\bibnamefont
  {Bluvstein}}, \bibinfo {author} {\bibfnamefont {S.~J.}\ \bibnamefont
  {Evered}}, \bibinfo {author} {\bibfnamefont {A.~A.}\ \bibnamefont {Geim}},
  \bibinfo {author} {\bibfnamefont {S.~H.}\ \bibnamefont {Li}}, \bibinfo
  {author} {\bibfnamefont {H.}~\bibnamefont {Zhou}}, \bibinfo {author}
  {\bibfnamefont {T.}~\bibnamefont {Manovitz}}, \bibinfo {author}
  {\bibfnamefont {S.}~\bibnamefont {Ebadi}}, \bibinfo {author} {\bibfnamefont
  {M.}~\bibnamefont {Cain}}, \bibinfo {author} {\bibfnamefont {M.}~\bibnamefont
  {Kalinowski}}, \bibinfo {author} {\bibfnamefont {D.}~\bibnamefont
  {Hangleiter}}, \bibinfo {author} {\bibfnamefont {J.~P.~B.}\ \bibnamefont
  {Ataides}}, \bibinfo {author} {\bibfnamefont {N.}~\bibnamefont {Maskara}},
  \bibinfo {author} {\bibfnamefont {I.}~\bibnamefont {Cong}}, \bibinfo {author}
  {\bibfnamefont {X.}~\bibnamefont {Gao}}, \bibinfo {author} {\bibfnamefont
  {P.~S.}\ \bibnamefont {Rodriguez}}, \bibinfo {author} {\bibfnamefont
  {T.}~\bibnamefont {Karolyshyn}}, \bibinfo {author} {\bibfnamefont
  {G.}~\bibnamefont {Semeghini}}, \bibinfo {author} {\bibfnamefont {M.~J.}\
  \bibnamefont {Gullans}}, \bibinfo {author} {\bibfnamefont {M.}~\bibnamefont
  {Greiner}}, \bibinfo {author} {\bibfnamefont {V.}~\bibnamefont {Vuletić}},\
  and\ \bibinfo {author} {\bibfnamefont {M.~D.}\ \bibnamefont {Lukin}},\ }\href
  {https://doi.org/10.1038/s41586-023-06927-3} {\bibfield  {journal} {\bibinfo
  {journal} {Nature}\ }\textbf {\bibinfo {volume} {626}},\ \bibinfo {pages}
  {58–65} (\bibinfo {year} {2023})}\BibitemShut {NoStop}%
\bibitem [{\citenamefont {da~Silva}\ \emph {et~al.}(2024)\citenamefont
  {da~Silva}, \citenamefont {Ryan-Anderson}, \citenamefont {Bello-Rivas},
  \citenamefont {Chernoguzov}, \citenamefont {Dreiling}, \citenamefont {Foltz},
  \citenamefont {Frachon}, \citenamefont {Gaebler}, \citenamefont {Gatterman},
  \citenamefont {Grans-Samuelsson}, \citenamefont {Hayes}, \citenamefont
  {Hewitt}, \citenamefont {Johansen}, \citenamefont {Lucchetti}, \citenamefont
  {Mills}, \citenamefont {Moses}, \citenamefont {Neyenhuis}, \citenamefont
  {Paz}, \citenamefont {Pino}, \citenamefont {Siegfried}, \citenamefont
  {Strabley}, \citenamefont {Sundaram}, \citenamefont {Tom}, \citenamefont
  {Wernli}, \citenamefont {Zanner}, \citenamefont {Stutz},\ and\ \citenamefont
  {Svore}}]{quantinuumMicrosoft2024}%
  \BibitemOpen
  \bibfield  {author} {\bibinfo {author} {\bibfnamefont {M.~P.}\ \bibnamefont
  {da~Silva}}, \bibinfo {author} {\bibfnamefont {C.}~\bibnamefont
  {Ryan-Anderson}}, \bibinfo {author} {\bibfnamefont {J.~M.}\ \bibnamefont
  {Bello-Rivas}}, \bibinfo {author} {\bibfnamefont {A.}~\bibnamefont
  {Chernoguzov}}, \bibinfo {author} {\bibfnamefont {J.~M.}\ \bibnamefont
  {Dreiling}}, \bibinfo {author} {\bibfnamefont {C.}~\bibnamefont {Foltz}},
  \bibinfo {author} {\bibfnamefont {F.}~\bibnamefont {Frachon}}, \bibinfo
  {author} {\bibfnamefont {J.~P.}\ \bibnamefont {Gaebler}}, \bibinfo {author}
  {\bibfnamefont {T.~M.}\ \bibnamefont {Gatterman}}, \bibinfo {author}
  {\bibfnamefont {L.}~\bibnamefont {Grans-Samuelsson}}, \bibinfo {author}
  {\bibfnamefont {D.}~\bibnamefont {Hayes}}, \bibinfo {author} {\bibfnamefont
  {N.}~\bibnamefont {Hewitt}}, \bibinfo {author} {\bibfnamefont
  {J.}~\bibnamefont {Johansen}}, \bibinfo {author} {\bibfnamefont
  {D.}~\bibnamefont {Lucchetti}}, \bibinfo {author} {\bibfnamefont
  {M.}~\bibnamefont {Mills}}, \bibinfo {author} {\bibfnamefont {S.~A.}\
  \bibnamefont {Moses}}, \bibinfo {author} {\bibfnamefont {B.}~\bibnamefont
  {Neyenhuis}}, \bibinfo {author} {\bibfnamefont {A.}~\bibnamefont {Paz}},
  \bibinfo {author} {\bibfnamefont {J.}~\bibnamefont {Pino}}, \bibinfo {author}
  {\bibfnamefont {P.}~\bibnamefont {Siegfried}}, \bibinfo {author}
  {\bibfnamefont {J.}~\bibnamefont {Strabley}}, \bibinfo {author}
  {\bibfnamefont {A.}~\bibnamefont {Sundaram}}, \bibinfo {author}
  {\bibfnamefont {D.}~\bibnamefont {Tom}}, \bibinfo {author} {\bibfnamefont
  {S.~J.}\ \bibnamefont {Wernli}}, \bibinfo {author} {\bibfnamefont
  {M.}~\bibnamefont {Zanner}}, \bibinfo {author} {\bibfnamefont {R.~P.}\
  \bibnamefont {Stutz}},\ and\ \bibinfo {author} {\bibfnamefont {K.~M.}\
  \bibnamefont {Svore}},\ }\bibfield  {journal} {\bibinfo  {journal}
  {arXiv:2404.02280}\ }\href {https://doi.org/10.48550/arXiv.2404.02280}
  {10.48550/arXiv.2404.02280} (\bibinfo {year} {2024})\BibitemShut {NoStop}%
\bibitem [{\citenamefont {Bharti}\ \emph {et~al.}(2022)\citenamefont {Bharti},
  \citenamefont {Cervera-Lierta}, \citenamefont {Kyaw}, \citenamefont {Haug},
  \citenamefont {Alperin-Lea}, \citenamefont {Anand}, \citenamefont {Degroote},
  \citenamefont {Heimonen}, \citenamefont {Kottmann}, \citenamefont {Menke},
  \citenamefont {Mok}, \citenamefont {Sim}, \citenamefont {Kwek},\ and\
  \citenamefont {Aspuru-Guzik}}]{NISQAlgo}%
  \BibitemOpen
  \bibfield  {author} {\bibinfo {author} {\bibfnamefont {K.}~\bibnamefont
  {Bharti}}, \bibinfo {author} {\bibfnamefont {A.}~\bibnamefont
  {Cervera-Lierta}}, \bibinfo {author} {\bibfnamefont {T.~H.}\ \bibnamefont
  {Kyaw}}, \bibinfo {author} {\bibfnamefont {T.}~\bibnamefont {Haug}}, \bibinfo
  {author} {\bibfnamefont {S.}~\bibnamefont {Alperin-Lea}}, \bibinfo {author}
  {\bibfnamefont {A.}~\bibnamefont {Anand}}, \bibinfo {author} {\bibfnamefont
  {M.}~\bibnamefont {Degroote}}, \bibinfo {author} {\bibfnamefont
  {H.}~\bibnamefont {Heimonen}}, \bibinfo {author} {\bibfnamefont {J.~S.}\
  \bibnamefont {Kottmann}}, \bibinfo {author} {\bibfnamefont {T.}~\bibnamefont
  {Menke}}, \bibinfo {author} {\bibfnamefont {W.-K.}\ \bibnamefont {Mok}},
  \bibinfo {author} {\bibfnamefont {S.}~\bibnamefont {Sim}}, \bibinfo {author}
  {\bibfnamefont {L.-C.}\ \bibnamefont {Kwek}},\ and\ \bibinfo {author}
  {\bibfnamefont {A.}~\bibnamefont {Aspuru-Guzik}},\ }\href
  {https://doi.org/10.1103/RevModPhys.94.015004} {\bibfield  {journal}
  {\bibinfo  {journal} {Rev. Mod. Phys.}\ }\textbf {\bibinfo {volume} {94}},\
  \bibinfo {pages} {015004} (\bibinfo {year} {2022})}\BibitemShut {NoStop}%
\bibitem [{\citenamefont {Deller}\ \emph {et~al.}(2023)\citenamefont {Deller},
  \citenamefont {Schmitt}, \citenamefont {Lewenstein}, \citenamefont {Lenk},
  \citenamefont {Federer}, \citenamefont {Jendrzejewski}, \citenamefont
  {Hauke},\ and\ \citenamefont {Kasper}}]{deller_quantum_2023}%
  \BibitemOpen
  \bibfield  {author} {\bibinfo {author} {\bibfnamefont {Y.}~\bibnamefont
  {Deller}}, \bibinfo {author} {\bibfnamefont {S.}~\bibnamefont {Schmitt}},
  \bibinfo {author} {\bibfnamefont {M.}~\bibnamefont {Lewenstein}}, \bibinfo
  {author} {\bibfnamefont {S.}~\bibnamefont {Lenk}}, \bibinfo {author}
  {\bibfnamefont {M.}~\bibnamefont {Federer}}, \bibinfo {author} {\bibfnamefont
  {F.}~\bibnamefont {Jendrzejewski}}, \bibinfo {author} {\bibfnamefont
  {P.}~\bibnamefont {Hauke}},\ and\ \bibinfo {author} {\bibfnamefont
  {V.}~\bibnamefont {Kasper}},\ }\href
  {https://doi.org/10.1103/PhysRevA.107.062410} {\bibfield  {journal} {\bibinfo
   {journal} {Physical Review A}\ }\textbf {\bibinfo {volume} {107}},\ \bibinfo
  {pages} {062410} (\bibinfo {year} {2023})}\BibitemShut {NoStop}%
\bibitem [{\citenamefont {{Yarkoni}}\ \emph {et~al.}(2022)\citenamefont
  {{Yarkoni}}, \citenamefont {{Raponi}}, \citenamefont {{B{\"a}ck}},\ and\
  \citenamefont {{Schmitt}}}]{yarkoni2021}%
  \BibitemOpen
  \bibfield  {author} {\bibinfo {author} {\bibfnamefont {S.}~\bibnamefont
  {{Yarkoni}}}, \bibinfo {author} {\bibfnamefont {E.}~\bibnamefont {{Raponi}}},
  \bibinfo {author} {\bibfnamefont {T.}~\bibnamefont {{B{\"a}ck}}},\ and\
  \bibinfo {author} {\bibfnamefont {S.}~\bibnamefont {{Schmitt}}},\ }\href
  {https://iopscience.iop.org/article/10.1088/1361-6633/ac8c54} {\bibfield
  {journal} {\bibinfo  {journal} {Reports on Progress in Physics}\ }\textbf
  {\bibinfo {volume} {85}},\ \bibinfo {pages} {104001} (\bibinfo {year}
  {2022})}\BibitemShut {NoStop}%
\bibitem [{\citenamefont {Abbas}\ \emph {et~al.}(2023)\citenamefont {Abbas},
  \citenamefont {Ambainis}, \citenamefont {Augustino}, \citenamefont
  {Bärtschi}, \citenamefont {Buhrman}, \citenamefont {Coffrin}, \citenamefont
  {Cortiana}, \citenamefont {Dunjko}, \citenamefont {Egger}, \citenamefont
  {Elmegreen}, \citenamefont {Franco}, \citenamefont {Fratini}, \citenamefont
  {Fuller}, \citenamefont {Gacon}, \citenamefont {Gonciulea}, \citenamefont
  {Gribling}, \citenamefont {Gupta}, \citenamefont {Hadfield}, \citenamefont
  {Heese}, \citenamefont {Kircher}, \citenamefont {Kleinert}, \citenamefont
  {Koch}, \citenamefont {Korpas}, \citenamefont {Lenk}, \citenamefont
  {Marecek}, \citenamefont {Markov}, \citenamefont {Mazzola}, \citenamefont
  {Mensa}, \citenamefont {Mohseni}, \citenamefont {Nannicini}, \citenamefont
  {O'Meara}, \citenamefont {Tapia}, \citenamefont {Pokutta}, \citenamefont
  {Proissl}, \citenamefont {Rebentrost}, \citenamefont {Sahin}, \citenamefont
  {Symons}, \citenamefont {Tornow}, \citenamefont {Valls}, \citenamefont
  {Woerner}, \citenamefont {Wolf-Bauwens}, \citenamefont {Yard}, \citenamefont
  {Yarkoni}, \citenamefont {Zechiel}, \citenamefont {Zhuk},\ and\ \citenamefont
  {Zoufal}}]{abbas_QuantumOptmization_2023}%
  \BibitemOpen
  \bibfield  {author} {\bibinfo {author} {\bibfnamefont {A.}~\bibnamefont
  {Abbas}}, \bibinfo {author} {\bibfnamefont {A.}~\bibnamefont {Ambainis}},
  \bibinfo {author} {\bibfnamefont {B.}~\bibnamefont {Augustino}}, \bibinfo
  {author} {\bibfnamefont {A.}~\bibnamefont {Bärtschi}}, \bibinfo {author}
  {\bibfnamefont {H.}~\bibnamefont {Buhrman}}, \bibinfo {author} {\bibfnamefont
  {C.}~\bibnamefont {Coffrin}}, \bibinfo {author} {\bibfnamefont
  {G.}~\bibnamefont {Cortiana}}, \bibinfo {author} {\bibfnamefont
  {V.}~\bibnamefont {Dunjko}}, \bibinfo {author} {\bibfnamefont {D.~J.}\
  \bibnamefont {Egger}}, \bibinfo {author} {\bibfnamefont {B.~G.}\ \bibnamefont
  {Elmegreen}}, \bibinfo {author} {\bibfnamefont {N.}~\bibnamefont {Franco}},
  \bibinfo {author} {\bibfnamefont {F.}~\bibnamefont {Fratini}}, \bibinfo
  {author} {\bibfnamefont {B.}~\bibnamefont {Fuller}}, \bibinfo {author}
  {\bibfnamefont {J.}~\bibnamefont {Gacon}}, \bibinfo {author} {\bibfnamefont
  {C.}~\bibnamefont {Gonciulea}}, \bibinfo {author} {\bibfnamefont
  {S.}~\bibnamefont {Gribling}}, \bibinfo {author} {\bibfnamefont
  {S.}~\bibnamefont {Gupta}}, \bibinfo {author} {\bibfnamefont
  {S.}~\bibnamefont {Hadfield}}, \bibinfo {author} {\bibfnamefont
  {R.}~\bibnamefont {Heese}}, \bibinfo {author} {\bibfnamefont
  {G.}~\bibnamefont {Kircher}}, \bibinfo {author} {\bibfnamefont
  {T.}~\bibnamefont {Kleinert}}, \bibinfo {author} {\bibfnamefont
  {T.}~\bibnamefont {Koch}}, \bibinfo {author} {\bibfnamefont {G.}~\bibnamefont
  {Korpas}}, \bibinfo {author} {\bibfnamefont {S.}~\bibnamefont {Lenk}},
  \bibinfo {author} {\bibfnamefont {J.}~\bibnamefont {Marecek}}, \bibinfo
  {author} {\bibfnamefont {V.}~\bibnamefont {Markov}}, \bibinfo {author}
  {\bibfnamefont {G.}~\bibnamefont {Mazzola}}, \bibinfo {author} {\bibfnamefont
  {S.}~\bibnamefont {Mensa}}, \bibinfo {author} {\bibfnamefont
  {N.}~\bibnamefont {Mohseni}}, \bibinfo {author} {\bibfnamefont
  {G.}~\bibnamefont {Nannicini}}, \bibinfo {author} {\bibfnamefont
  {C.}~\bibnamefont {O'Meara}}, \bibinfo {author} {\bibfnamefont {E.~P.}\
  \bibnamefont {Tapia}}, \bibinfo {author} {\bibfnamefont {S.}~\bibnamefont
  {Pokutta}}, \bibinfo {author} {\bibfnamefont {M.}~\bibnamefont {Proissl}},
  \bibinfo {author} {\bibfnamefont {P.}~\bibnamefont {Rebentrost}}, \bibinfo
  {author} {\bibfnamefont {E.}~\bibnamefont {Sahin}}, \bibinfo {author}
  {\bibfnamefont {B.~C.~B.}\ \bibnamefont {Symons}}, \bibinfo {author}
  {\bibfnamefont {S.}~\bibnamefont {Tornow}}, \bibinfo {author} {\bibfnamefont
  {V.}~\bibnamefont {Valls}}, \bibinfo {author} {\bibfnamefont
  {S.}~\bibnamefont {Woerner}}, \bibinfo {author} {\bibfnamefont {M.~L.}\
  \bibnamefont {Wolf-Bauwens}}, \bibinfo {author} {\bibfnamefont
  {J.}~\bibnamefont {Yard}}, \bibinfo {author} {\bibfnamefont {S.}~\bibnamefont
  {Yarkoni}}, \bibinfo {author} {\bibfnamefont {D.}~\bibnamefont {Zechiel}},
  \bibinfo {author} {\bibfnamefont {S.}~\bibnamefont {Zhuk}},\ and\ \bibinfo
  {author} {\bibfnamefont {C.}~\bibnamefont {Zoufal}},\ }\bibfield  {journal}
  {\bibinfo  {journal} {arXiv:2312.02279}\ }\href
  {https://doi.org/10.48550/arXiv.2312.02279} {10.48550/arXiv.2312.02279}
  (\bibinfo {year} {2023})\BibitemShut {NoStop}%
\bibitem [{\citenamefont {Veshchezerova}\ \emph {et~al.}(2023)\citenamefont
  {Veshchezerova}, \citenamefont {Somov}, \citenamefont {Bertsche},
  \citenamefont {Limmer}, \citenamefont {Schmitt}, \citenamefont
  {Perelshtein},\ and\ \citenamefont {Joshi~Tripathi}}]{tq_EVcharging2023}%
  \BibitemOpen
  \bibfield  {author} {\bibinfo {author} {\bibfnamefont {M.}~\bibnamefont
  {Veshchezerova}}, \bibinfo {author} {\bibfnamefont {M.}~\bibnamefont
  {Somov}}, \bibinfo {author} {\bibfnamefont {D.}~\bibnamefont {Bertsche}},
  \bibinfo {author} {\bibfnamefont {S.}~\bibnamefont {Limmer}}, \bibinfo
  {author} {\bibfnamefont {S.}~\bibnamefont {Schmitt}}, \bibinfo {author}
  {\bibfnamefont {M.}~\bibnamefont {Perelshtein}},\ and\ \bibinfo {author}
  {\bibfnamefont {A.}~\bibnamefont {Joshi~Tripathi}},\ }in\ \href
  {https://doi.org/10.1109/QCE57702.2023.00078} {\emph {\bibinfo {booktitle}
  {2023 IEEE International Conference on Quantum Computing and Engineering
  (QCE)}}},\ Vol.~\bibinfo {volume} {01}\ (\bibinfo {year} {2023})\ pp.\
  \bibinfo {pages} {636--641}\BibitemShut {NoStop}%
\bibitem [{\citenamefont {{Dalyac, Constantin}}\ \emph
  {et~al.}(2021)\citenamefont {{Dalyac, Constantin}}, \citenamefont {{Henriet,
  Lo\"{\i}c}}, \citenamefont {{Jeandel, Emmanuel}}, \citenamefont {{Lechner,
  Wolfgang}}, \citenamefont {{Perdrix, Simon}}, \citenamefont {{Porcheron,
  Marc}},\ and\ \citenamefont {{Veshchezerova,
  Margarita}}}]{dalyacSmartChargingQAOA2021}%
  \BibitemOpen
  \bibfield  {author} {\bibinfo {author} {\bibnamefont {{Dalyac, Constantin}}},
  \bibinfo {author} {\bibnamefont {{Henriet, Lo\"{\i}c}}}, \bibinfo {author}
  {\bibnamefont {{Jeandel, Emmanuel}}}, \bibinfo {author} {\bibnamefont
  {{Lechner, Wolfgang}}}, \bibinfo {author} {\bibnamefont {{Perdrix, Simon}}},
  \bibinfo {author} {\bibnamefont {{Porcheron, Marc}}},\ and\ \bibinfo {author}
  {\bibnamefont {{Veshchezerova, Margarita}}},\ }\href
  {https://doi.org/10.1140/epjqt/s40507-021-00100-3} {\bibfield  {journal}
  {\bibinfo  {journal} {EPJ Quantum Technol.}\ }\textbf {\bibinfo {volume}
  {8}},\ \bibinfo {pages} {12} (\bibinfo {year} {2021})}\BibitemShut {NoStop}%
\bibitem [{\citenamefont {Farhi}\ \emph {et~al.}(2014)\citenamefont {Farhi},
  \citenamefont {Goldstone},\ and\ \citenamefont {Gutmann}}]{farhi2014quantum}%
  \BibitemOpen
  \bibfield  {author} {\bibinfo {author} {\bibfnamefont {E.}~\bibnamefont
  {Farhi}}, \bibinfo {author} {\bibfnamefont {J.}~\bibnamefont {Goldstone}},\
  and\ \bibinfo {author} {\bibfnamefont {S.}~\bibnamefont {Gutmann}},\
  }\bibfield  {journal} {\bibinfo  {journal} {arXiv:1411.4028}\ }\href
  {https://doi.org/10.48550/arXiv.1411.4028} {10.48550/arXiv.1411.4028}
  (\bibinfo {year} {2014})\BibitemShut {NoStop}%
\bibitem [{\citenamefont {Blekos}\ \emph {et~al.}(2024)\citenamefont {Blekos},
  \citenamefont {Brand}, \citenamefont {Ceschini}, \citenamefont {Chou},
  \citenamefont {Li}, \citenamefont {Pandya},\ and\ \citenamefont
  {Summer}}]{blekosQAOA_Review}%
  \BibitemOpen
  \bibfield  {author} {\bibinfo {author} {\bibfnamefont {K.}~\bibnamefont
  {Blekos}}, \bibinfo {author} {\bibfnamefont {D.}~\bibnamefont {Brand}},
  \bibinfo {author} {\bibfnamefont {A.}~\bibnamefont {Ceschini}}, \bibinfo
  {author} {\bibfnamefont {C.-H.}\ \bibnamefont {Chou}}, \bibinfo {author}
  {\bibfnamefont {R.-H.}\ \bibnamefont {Li}}, \bibinfo {author} {\bibfnamefont
  {K.}~\bibnamefont {Pandya}},\ and\ \bibinfo {author} {\bibfnamefont
  {A.}~\bibnamefont {Summer}},\ }\href
  {https://doi.org/https://doi.org/10.1016/j.physrep.2024.03.002} {\bibfield
  {journal} {\bibinfo  {journal} {Physics Reports}\ }\textbf {\bibinfo {volume}
  {1068}},\ \bibinfo {pages} {1} (\bibinfo {year} {2024})},\ \bibinfo {note} {a
  review on Quantum Approximate Optimization Algorithm and its
  variants}\BibitemShut {NoStop}%
\bibitem [{\citenamefont {Farhi}\ \emph {et~al.}(2015)\citenamefont {Farhi},
  \citenamefont {Goldstone},\ and\ \citenamefont {Gutmann}}]{farhi2015quantum}%
  \BibitemOpen
  \bibfield  {author} {\bibinfo {author} {\bibfnamefont {E.}~\bibnamefont
  {Farhi}}, \bibinfo {author} {\bibfnamefont {J.}~\bibnamefont {Goldstone}},\
  and\ \bibinfo {author} {\bibfnamefont {S.}~\bibnamefont {Gutmann}},\
  }\bibfield  {journal} {\bibinfo  {journal} {arXiv:1412.6062}\ }\href
  {https://doi.org/10.48550/arXiv.1412.6062} {10.48550/arXiv.1412.6062}
  (\bibinfo {year} {2015})\BibitemShut {NoStop}%
\bibitem [{\citenamefont {Farhi}\ and\ \citenamefont
  {Harrow}(2019)}]{farhi2019quantum}%
  \BibitemOpen
  \bibfield  {author} {\bibinfo {author} {\bibfnamefont {E.}~\bibnamefont
  {Farhi}}\ and\ \bibinfo {author} {\bibfnamefont {A.~W.}\ \bibnamefont
  {Harrow}},\ }\bibfield  {journal} {\bibinfo  {journal} {arXiv:1602.07674}\
  }\href {https://doi.org/10.48550/arXiv.1602.07674}
  {10.48550/arXiv.1602.07674} (\bibinfo {year} {2019})\BibitemShut {NoStop}%
\bibitem [{\citenamefont {Sharma}\ and\ \citenamefont
  {Kumar}(2022)}]{SharmaMOOreview2022}%
  \BibitemOpen
  \bibfield  {author} {\bibinfo {author} {\bibfnamefont {S.}~\bibnamefont
  {Sharma}}\ and\ \bibinfo {author} {\bibfnamefont {V.~A.}\ \bibnamefont
  {Kumar}},\ }\href {https://doi.org/10.1007/s11831-022-09778-9} {\bibfield
  {journal} {\bibinfo  {journal} {Arch Computat Methods Eng}\ }\textbf
  {\bibinfo {volume} {29}},\ \bibinfo {pages} {5605} (\bibinfo {year}
  {2022})}\BibitemShut {NoStop}%
\bibitem [{\citenamefont {Pereira}\ \emph {et~al.}(2022)\citenamefont
  {Pereira}, \citenamefont {Oliver}, \citenamefont {Francisco}, \citenamefont
  {Jr},\ and\ \citenamefont {Gomes}}]{pereiraMOOreview2022}%
  \BibitemOpen
  \bibfield  {author} {\bibinfo {author} {\bibfnamefont {J.~L.~J.}\
  \bibnamefont {Pereira}}, \bibinfo {author} {\bibfnamefont {G.~A.}\
  \bibnamefont {Oliver}}, \bibinfo {author} {\bibfnamefont {M.~B.}\
  \bibnamefont {Francisco}}, \bibinfo {author} {\bibfnamefont {S.~S.~C.}\
  \bibnamefont {Jr}},\ and\ \bibinfo {author} {\bibfnamefont {G.~F.}\
  \bibnamefont {Gomes}},\ }\href {https://doi.org/10.1007/s11831-021-09663-x}
  {\bibfield  {journal} {\bibinfo  {journal} {Arch Computat Methods Eng}\
  }\textbf {\bibinfo {volume} {29}},\ \bibinfo {pages} {2285} (\bibinfo {year}
  {2022})}\BibitemShut {NoStop}%
\bibitem [{\citenamefont {Chancellor}(2019)}]{ChancellorDomainWall2019}%
  \BibitemOpen
  \bibfield  {author} {\bibinfo {author} {\bibfnamefont {N.}~\bibnamefont
  {Chancellor}},\ }\href {https://doi.org/10.1088/2058-9565/ab33c2} {\bibfield
  {journal} {\bibinfo  {journal} {Quantum Science and Technology}\ }\textbf
  {\bibinfo {volume} {4}},\ \bibinfo {pages} {045004} (\bibinfo {year}
  {2019})}\BibitemShut {NoStop}%
\bibitem [{\citenamefont {Sawaya}\ \emph {et~al.}(2023)\citenamefont {Sawaya},
  \citenamefont {Schmitz},\ and\ \citenamefont
  {Hadfield}}]{Sawaya2023encodingtradeoffs}%
  \BibitemOpen
  \bibfield  {author} {\bibinfo {author} {\bibfnamefont {N.~P.}\ \bibnamefont
  {Sawaya}}, \bibinfo {author} {\bibfnamefont {A.~T.}\ \bibnamefont
  {Schmitz}},\ and\ \bibinfo {author} {\bibfnamefont {S.}~\bibnamefont
  {Hadfield}},\ }\href {https://doi.org/10.22331/q-2023-09-14-1111} {\bibfield
  {journal} {\bibinfo  {journal} {{Quantum}}\ }\textbf {\bibinfo {volume}
  {7}},\ \bibinfo {pages} {1111} (\bibinfo {year} {2023})}\BibitemShut
  {NoStop}%
\bibitem [{\citenamefont {Wang}\ \emph {et~al.}(2020)\citenamefont {Wang},
  \citenamefont {Hu}, \citenamefont {Sanders},\ and\ \citenamefont
  {Kais}}]{wang_qudits_2020}%
  \BibitemOpen
  \bibfield  {author} {\bibinfo {author} {\bibfnamefont {Y.}~\bibnamefont
  {Wang}}, \bibinfo {author} {\bibfnamefont {Z.}~\bibnamefont {Hu}}, \bibinfo
  {author} {\bibfnamefont {B.~C.}\ \bibnamefont {Sanders}},\ and\ \bibinfo
  {author} {\bibfnamefont {S.}~\bibnamefont {Kais}},\ }\href
  {https://doi.org/10.3389/fphy.2020.589504} {\bibfield  {journal} {\bibinfo
  {journal} {Frontiers in Physics}\ }\textbf {\bibinfo {volume} {8}},\ \bibinfo
  {pages} {589504} (\bibinfo {year} {2020})}\BibitemShut {NoStop}%
\bibitem [{\citenamefont {Lanyon}\ \emph {et~al.}(2009)\citenamefont {Lanyon},
  \citenamefont {Barbieri}, \citenamefont {Almeida}, \citenamefont {Jennewein},
  \citenamefont {Ralph}, \citenamefont {Resch}, \citenamefont {Pryde},
  \citenamefont {O’Brien}, \citenamefont {Gilchrist},\ and\ \citenamefont
  {White}}]{quditsSimplify2008}%
  \BibitemOpen
  \bibfield  {author} {\bibinfo {author} {\bibfnamefont {B.~P.}\ \bibnamefont
  {Lanyon}}, \bibinfo {author} {\bibfnamefont {M.}~\bibnamefont {Barbieri}},
  \bibinfo {author} {\bibfnamefont {M.~P.}\ \bibnamefont {Almeida}}, \bibinfo
  {author} {\bibfnamefont {T.}~\bibnamefont {Jennewein}}, \bibinfo {author}
  {\bibfnamefont {T.~C.}\ \bibnamefont {Ralph}}, \bibinfo {author}
  {\bibfnamefont {K.~J.}\ \bibnamefont {Resch}}, \bibinfo {author}
  {\bibfnamefont {G.~J.}\ \bibnamefont {Pryde}}, \bibinfo {author}
  {\bibfnamefont {J.~L.}\ \bibnamefont {O’Brien}}, \bibinfo {author}
  {\bibfnamefont {A.}~\bibnamefont {Gilchrist}},\ and\ \bibinfo {author}
  {\bibfnamefont {A.~G.}\ \bibnamefont {White}},\ }\href
  {https://doi.org/10.1038/nphys1150} {\bibfield  {journal} {\bibinfo
  {journal} {Nature Physics}\ }\textbf {\bibinfo {volume} {5}},\ \bibinfo
  {pages} {134–140} (\bibinfo {year} {2009})}\BibitemShut {NoStop}%
\bibitem [{\citenamefont {Gao}\ \emph {et~al.}(2023)\citenamefont {Gao},
  \citenamefont {Appel}, \citenamefont {Friis}, \citenamefont {Ringbauer},\
  and\ \citenamefont {Huber}}]{Gao2023roleofentanglement}%
  \BibitemOpen
  \bibfield  {author} {\bibinfo {author} {\bibfnamefont {X.}~\bibnamefont
  {Gao}}, \bibinfo {author} {\bibfnamefont {P.}~\bibnamefont {Appel}}, \bibinfo
  {author} {\bibfnamefont {N.}~\bibnamefont {Friis}}, \bibinfo {author}
  {\bibfnamefont {M.}~\bibnamefont {Ringbauer}},\ and\ \bibinfo {author}
  {\bibfnamefont {M.}~\bibnamefont {Huber}},\ }\href
  {https://doi.org/10.22331/q-2023-10-16-1141} {\bibfield  {journal} {\bibinfo
  {journal} {{Quantum}}\ }\textbf {\bibinfo {volume} {7}},\ \bibinfo {pages}
  {1141} (\bibinfo {year} {2023})}\BibitemShut {NoStop}%
\bibitem [{\citenamefont {Duclos-Cianci}\ and\ \citenamefont
  {Poulin}(2013)}]{qudit_errorCorr2013}%
  \BibitemOpen
  \bibfield  {author} {\bibinfo {author} {\bibfnamefont {G.}~\bibnamefont
  {Duclos-Cianci}}\ and\ \bibinfo {author} {\bibfnamefont {D.}~\bibnamefont
  {Poulin}},\ }\href {https://doi.org/10.1103/PhysRevA.87.062338} {\bibfield
  {journal} {\bibinfo  {journal} {Phys. Rev. A}\ }\textbf {\bibinfo {volume}
  {87}},\ \bibinfo {pages} {062338} (\bibinfo {year} {2013})}\BibitemShut
  {NoStop}%
\bibitem [{\citenamefont {Watson}\ \emph {et~al.}(2015)\citenamefont {Watson},
  \citenamefont {Anwar},\ and\ \citenamefont {Browne}}]{qudit_errorCorr2015}%
  \BibitemOpen
  \bibfield  {author} {\bibinfo {author} {\bibfnamefont {F.~H.~E.}\
  \bibnamefont {Watson}}, \bibinfo {author} {\bibfnamefont {H.}~\bibnamefont
  {Anwar}},\ and\ \bibinfo {author} {\bibfnamefont {D.~E.}\ \bibnamefont
  {Browne}},\ }\href {https://doi.org/10.1103/PhysRevA.92.032309} {\bibfield
  {journal} {\bibinfo  {journal} {Phys. Rev. A}\ }\textbf {\bibinfo {volume}
  {92}},\ \bibinfo {pages} {032309} (\bibinfo {year} {2015})}\BibitemShut
  {NoStop}%
\bibitem [{\citenamefont {Gedik}\ \emph {et~al.}(2015)\citenamefont {Gedik},
  \citenamefont {Silva}, \citenamefont {Çakmak}, \citenamefont {Karpat},
  \citenamefont {Vidoto}, \citenamefont {Soares-Pinto}, \citenamefont
  {deAzevedo},\ and\ \citenamefont {Fanchini}}]{gedik_qudit_speedup2015}%
  \BibitemOpen
  \bibfield  {author} {\bibinfo {author} {\bibfnamefont {Z.}~\bibnamefont
  {Gedik}}, \bibinfo {author} {\bibfnamefont {I.}~\bibnamefont {Silva}},
  \bibinfo {author} {\bibfnamefont {B.}~\bibnamefont {Çakmak}}, \bibinfo
  {author} {\bibfnamefont {G.}~\bibnamefont {Karpat}}, \bibinfo {author}
  {\bibfnamefont {E.~L.~G.}\ \bibnamefont {Vidoto}}, \bibinfo {author}
  {\bibfnamefont {D.~O.}\ \bibnamefont {Soares-Pinto}}, \bibinfo {author}
  {\bibfnamefont {E.~R.}\ \bibnamefont {deAzevedo}},\ and\ \bibinfo {author}
  {\bibfnamefont {F.~F.}\ \bibnamefont {Fanchini}},\ }\href
  {https://doi.org/10.1038/srep14671} {\bibfield  {journal} {\bibinfo
  {journal} {Sci Rep}\ }\textbf {\bibinfo {volume} {5}},\ \bibinfo {pages}
  {14671} (\bibinfo {year} {2015})}\BibitemShut {NoStop}%
\bibitem [{\citenamefont {Bottrill}\ \emph {et~al.}(2023)\citenamefont
  {Bottrill}, \citenamefont {Pandey},\ and\ \citenamefont
  {Di~Matteo}}]{bottrillQutrit2023}%
  \BibitemOpen
  \bibfield  {author} {\bibinfo {author} {\bibfnamefont {G.}~\bibnamefont
  {Bottrill}}, \bibinfo {author} {\bibfnamefont {M.}~\bibnamefont {Pandey}},\
  and\ \bibinfo {author} {\bibfnamefont {O.}~\bibnamefont {Di~Matteo}},\ }in\
  \href {https://doi.org/10.1109/QCE57702.2023.00028} {\emph {\bibinfo
  {booktitle} {2023 IEEE International Conference on Quantum Computing and
  Engineering (QCE)}}},\ Vol.~\bibinfo {volume} {01}\ (\bibinfo {year} {2023})\
  pp.\ \bibinfo {pages} {177--183}\BibitemShut {NoStop}%
\bibitem [{\citenamefont {Bravyi}\ \emph {et~al.}(2022)\citenamefont {Bravyi},
  \citenamefont {Kliesch}, \citenamefont {Koenig},\ and\ \citenamefont
  {Tang}}]{bravyi2022}%
  \BibitemOpen
  \bibfield  {author} {\bibinfo {author} {\bibfnamefont {S.}~\bibnamefont
  {Bravyi}}, \bibinfo {author} {\bibfnamefont {A.}~\bibnamefont {Kliesch}},
  \bibinfo {author} {\bibfnamefont {R.}~\bibnamefont {Koenig}},\ and\ \bibinfo
  {author} {\bibfnamefont {E.}~\bibnamefont {Tang}},\ }\href@noop {} {\bibfield
   {journal} {\bibinfo  {journal} {Quantum}\ }\textbf {\bibinfo {volume} {6}},\
  \bibinfo {pages} {678} (\bibinfo {year} {2022})}\BibitemShut {NoStop}%
\bibitem [{\citenamefont {Wach}\ \emph {et~al.}(2023)\citenamefont {Wach},
  \citenamefont {Rudolph}, \citenamefont {Jendrzejewski},\ and\ \citenamefont
  {Schmitt}}]{wachDRULQudit2022}%
  \BibitemOpen
  \bibfield  {author} {\bibinfo {author} {\bibfnamefont {N.~L.}\ \bibnamefont
  {Wach}}, \bibinfo {author} {\bibfnamefont {M.~S.}\ \bibnamefont {Rudolph}},
  \bibinfo {author} {\bibfnamefont {F.}~\bibnamefont {Jendrzejewski}},\ and\
  \bibinfo {author} {\bibfnamefont {S.}~\bibnamefont {Schmitt}},\ }\href
  {https://doi.org/10.1007/s42484-023-00125-0} {\bibfield  {journal} {\bibinfo
  {journal} {Quantum Mach. Intell.}\ }\textbf {\bibinfo {volume} {5}},\
  \bibinfo {pages} {36} (\bibinfo {year} {2023})}\BibitemShut {NoStop}%
\bibitem [{\citenamefont {Roca-Jerat}\ \emph {et~al.}(2023)\citenamefont
  {Roca-Jerat}, \citenamefont {Román-Roche},\ and\ \citenamefont
  {Zueco}}]{rocaJeratquditML2023}%
  \BibitemOpen
  \bibfield  {author} {\bibinfo {author} {\bibfnamefont {S.}~\bibnamefont
  {Roca-Jerat}}, \bibinfo {author} {\bibfnamefont {J.}~\bibnamefont
  {Román-Roche}},\ and\ \bibinfo {author} {\bibfnamefont {D.}~\bibnamefont
  {Zueco}},\ }\bibfield  {journal} {\bibinfo  {journal} {arXiv:2308.16230}\
  }\href {https://doi.org/10.48550/arXiv.2308.16230}
  {10.48550/arXiv.2308.16230} (\bibinfo {year} {2023})\BibitemShut {NoStop}%
\bibitem [{\citenamefont {Useche}\ \emph {et~al.}(2022)\citenamefont {Useche},
  \citenamefont {Giraldo-Carvajal}, \citenamefont {Zuluaga-Bucheli},
  \citenamefont {Jaramillo-Villegas},\ and\ \citenamefont
  {González}}]{qudit_KDE_2022}%
  \BibitemOpen
  \bibfield  {author} {\bibinfo {author} {\bibfnamefont {D.~H.}\ \bibnamefont
  {Useche}}, \bibinfo {author} {\bibfnamefont {A.}~\bibnamefont
  {Giraldo-Carvajal}}, \bibinfo {author} {\bibfnamefont {H.~M.}\ \bibnamefont
  {Zuluaga-Bucheli}}, \bibinfo {author} {\bibfnamefont {J.~A.}\ \bibnamefont
  {Jaramillo-Villegas}},\ and\ \bibinfo {author} {\bibfnamefont {F.~A.}\
  \bibnamefont {González}},\ }\href
  {https://doi.org/10.1007/s11128-021-03363-y} {\bibfield  {journal} {\bibinfo
  {journal} {Quantum Inf Process}\ }\textbf {\bibinfo {volume} {21}},\ \bibinfo
  {pages} {12} (\bibinfo {year} {2022})}\BibitemShut {NoStop}%
\bibitem [{\citenamefont {Garcia~de Andoin}\ \emph {et~al.}(2023)\citenamefont
  {Garcia~de Andoin}, \citenamefont {Bottarelli}, \citenamefont {Schmitt},
  \citenamefont {Oregi}, \citenamefont {Hauke},\ and\ \citenamefont
  {Sanz}}]{mikelEVqudits2023}%
  \BibitemOpen
  \bibfield  {author} {\bibinfo {author} {\bibfnamefont {M.}~\bibnamefont
  {Garcia~de Andoin}}, \bibinfo {author} {\bibfnamefont {A.}~\bibnamefont
  {Bottarelli}}, \bibinfo {author} {\bibfnamefont {S.}~\bibnamefont {Schmitt}},
  \bibinfo {author} {\bibfnamefont {I.}~\bibnamefont {Oregi}}, \bibinfo
  {author} {\bibfnamefont {P.}~\bibnamefont {Hauke}},\ and\ \bibinfo {author}
  {\bibfnamefont {M.}~\bibnamefont {Sanz}},\ }\bibfield  {journal} {\bibinfo
  {journal} {arXiv:2306.04414}\ }\href
  {https://doi.org/10.48550/arXiv.2306.04414} {10.48550/arXiv.2306.04414}
  (\bibinfo {year} {2023})\BibitemShut {NoStop}%
\bibitem [{\citenamefont {Vargas-Calder\'{o}n}\ \emph
  {et~al.}(2021)\citenamefont {Vargas-Calder\'{o}n}, \citenamefont {Parra-A.},
  \citenamefont {Vinck-Posada},\ and\ \citenamefont
  {Gonz\'{a}lez}}]{vargas_TSP_qudits2021}%
  \BibitemOpen
  \bibfield  {author} {\bibinfo {author} {\bibfnamefont {V.}~\bibnamefont
  {Vargas-Calder\'{o}n}}, \bibinfo {author} {\bibfnamefont {N.}~\bibnamefont
  {Parra-A.}}, \bibinfo {author} {\bibfnamefont {H.}~\bibnamefont
  {Vinck-Posada}},\ and\ \bibinfo {author} {\bibfnamefont {F.~A.}\ \bibnamefont
  {Gonz\'{a}lez}},\ }\href {https://doi.org/10.7566/JPSJ.90.114002} {\bibfield
  {journal} {\bibinfo  {journal} {Journal of the Physical Society of Japan}\
  }\textbf {\bibinfo {volume} {90}},\ \bibinfo {pages} {114002} (\bibinfo
  {year} {2021})}\BibitemShut {NoStop}%
\bibitem [{\citenamefont {Bottarelli}\ \emph {et~al.}(2024)\citenamefont
  {Bottarelli}, \citenamefont {Schmitt},\ and\ \citenamefont
  {Hauke}}]{bottarelliConstraints2024}%
  \BibitemOpen
  \bibfield  {author} {\bibinfo {author} {\bibfnamefont {A.}~\bibnamefont
  {Bottarelli}}, \bibinfo {author} {\bibfnamefont {S.}~\bibnamefont
  {Schmitt}},\ and\ \bibinfo {author} {\bibfnamefont {P.}~\bibnamefont
  {Hauke}},\ }\bibfield  {journal} {\bibinfo  {journal} {arXiv:2410.07674}\
  }\href {https://doi.org/10.48550/arXiv.2410.07674}
  {10.48550/arXiv.2410.07674} (\bibinfo {year} {2024})\BibitemShut {NoStop}%
\bibitem [{\citenamefont {Ringbauer}\ \emph {et~al.}(2022)\citenamefont
  {Ringbauer}, \citenamefont {Meth}, \citenamefont {Postler}, \citenamefont
  {Stricker}, \citenamefont {Blatt}, \citenamefont {Schindler},\ and\
  \citenamefont {Monz}}]{ringbauerIonQudits2022}%
  \BibitemOpen
  \bibfield  {author} {\bibinfo {author} {\bibfnamefont {M.}~\bibnamefont
  {Ringbauer}}, \bibinfo {author} {\bibfnamefont {M.}~\bibnamefont {Meth}},
  \bibinfo {author} {\bibfnamefont {L.}~\bibnamefont {Postler}}, \bibinfo
  {author} {\bibfnamefont {R.}~\bibnamefont {Stricker}}, \bibinfo {author}
  {\bibfnamefont {R.}~\bibnamefont {Blatt}}, \bibinfo {author} {\bibfnamefont
  {P.}~\bibnamefont {Schindler}},\ and\ \bibinfo {author} {\bibfnamefont
  {T.}~\bibnamefont {Monz}},\ }\href
  {https://doi.org/10.1038/s41567-022-01658-0} {\bibfield  {journal} {\bibinfo
  {journal} {Nat. Phys.}\ }\textbf {\bibinfo {volume} {18}},\ \bibinfo {pages}
  {1053–1057} (\bibinfo {year} {2022})}\BibitemShut {NoStop}%
\bibitem [{\citenamefont {Hrmo}\ \emph {et~al.}(2023)\citenamefont {Hrmo},
  \citenamefont {Wilhelm}, \citenamefont {Gerster}, \citenamefont {van Mourik},
  \citenamefont {Huber}, \citenamefont {Blatt}, \citenamefont {Schindler},
  \citenamefont {Monz},\ and\ \citenamefont
  {Ringbauer}}]{ringbauer_entanglement2023}%
  \BibitemOpen
  \bibfield  {author} {\bibinfo {author} {\bibfnamefont {P.}~\bibnamefont
  {Hrmo}}, \bibinfo {author} {\bibfnamefont {B.}~\bibnamefont {Wilhelm}},
  \bibinfo {author} {\bibfnamefont {L.}~\bibnamefont {Gerster}}, \bibinfo
  {author} {\bibfnamefont {M.~W.}\ \bibnamefont {van Mourik}}, \bibinfo
  {author} {\bibfnamefont {M.}~\bibnamefont {Huber}}, \bibinfo {author}
  {\bibfnamefont {R.}~\bibnamefont {Blatt}}, \bibinfo {author} {\bibfnamefont
  {P.}~\bibnamefont {Schindler}}, \bibinfo {author} {\bibfnamefont
  {T.}~\bibnamefont {Monz}},\ and\ \bibinfo {author} {\bibfnamefont
  {M.}~\bibnamefont {Ringbauer}},\ }\href
  {https://doi.org/10.1038/s41467-023-37375-2} {\bibfield  {journal} {\bibinfo
  {journal} {Nature Communications}\ }\textbf {\bibinfo {volume} {14}},\
  \bibinfo {pages} {2242} (\bibinfo {year} {2023})}\BibitemShut {NoStop}%
\bibitem [{\citenamefont {Chi}\ \emph {et~al.}(2022)\citenamefont {Chi},
  \citenamefont {Huang}, \citenamefont {Zhang}, \citenamefont {Mao},
  \citenamefont {Zhou}, \citenamefont {Chen}, \citenamefont {Zhai},
  \citenamefont {Bao}, \citenamefont {Dai}, \citenamefont {Yuan}, \citenamefont
  {Zhang}, \citenamefont {Dai}, \citenamefont {Tang}, \citenamefont {Yang},
  \citenamefont {Li}, \citenamefont {Ding}, \citenamefont {Oxenløwe},
  \citenamefont {Thompson}, \citenamefont {O’Brien}, \citenamefont {Li},
  \citenamefont {Gong},\ and\ \citenamefont {Wang}}]{chi_quditprocessor2022}%
  \BibitemOpen
  \bibfield  {author} {\bibinfo {author} {\bibfnamefont {Y.}~\bibnamefont
  {Chi}}, \bibinfo {author} {\bibfnamefont {J.}~\bibnamefont {Huang}}, \bibinfo
  {author} {\bibfnamefont {Z.}~\bibnamefont {Zhang}}, \bibinfo {author}
  {\bibfnamefont {J.}~\bibnamefont {Mao}}, \bibinfo {author} {\bibfnamefont
  {Z.}~\bibnamefont {Zhou}}, \bibinfo {author} {\bibfnamefont {X.}~\bibnamefont
  {Chen}}, \bibinfo {author} {\bibfnamefont {C.}~\bibnamefont {Zhai}}, \bibinfo
  {author} {\bibfnamefont {J.}~\bibnamefont {Bao}}, \bibinfo {author}
  {\bibfnamefont {T.}~\bibnamefont {Dai}}, \bibinfo {author} {\bibfnamefont
  {H.}~\bibnamefont {Yuan}}, \bibinfo {author} {\bibfnamefont {M.}~\bibnamefont
  {Zhang}}, \bibinfo {author} {\bibfnamefont {D.}~\bibnamefont {Dai}}, \bibinfo
  {author} {\bibfnamefont {B.}~\bibnamefont {Tang}}, \bibinfo {author}
  {\bibfnamefont {Y.}~\bibnamefont {Yang}}, \bibinfo {author} {\bibfnamefont
  {Z.}~\bibnamefont {Li}}, \bibinfo {author} {\bibfnamefont {Y.}~\bibnamefont
  {Ding}}, \bibinfo {author} {\bibfnamefont {L.~K.}\ \bibnamefont {Oxenløwe}},
  \bibinfo {author} {\bibfnamefont {M.~G.}\ \bibnamefont {Thompson}}, \bibinfo
  {author} {\bibfnamefont {J.~L.}\ \bibnamefont {O’Brien}}, \bibinfo {author}
  {\bibfnamefont {Y.}~\bibnamefont {Li}}, \bibinfo {author} {\bibfnamefont
  {Q.}~\bibnamefont {Gong}},\ and\ \bibinfo {author} {\bibfnamefont
  {J.}~\bibnamefont {Wang}},\ }\href
  {https://doi.org/10.1038/s41467-022-28767-x} {\bibfield  {journal} {\bibinfo
  {journal} {Nature Communications}\ }\textbf {\bibinfo {volume} {13}},\
  \bibinfo {pages} {1166} (\bibinfo {year} {2022})}\BibitemShut {NoStop}%
\bibitem [{\citenamefont {Blok}\ \emph {et~al.}(2021)\citenamefont {Blok},
  \citenamefont {Ramasesh}, \citenamefont {Schuster}, \citenamefont {O'Brien},
  \citenamefont {Kreikebaum}, \citenamefont {Dahlen}, \citenamefont {Morvan},
  \citenamefont {Yoshida}, \citenamefont {Yao},\ and\ \citenamefont
  {Siddiqi}}]{qutrit_experiement2021}%
  \BibitemOpen
  \bibfield  {author} {\bibinfo {author} {\bibfnamefont {M.~S.}\ \bibnamefont
  {Blok}}, \bibinfo {author} {\bibfnamefont {V.~V.}\ \bibnamefont {Ramasesh}},
  \bibinfo {author} {\bibfnamefont {T.}~\bibnamefont {Schuster}}, \bibinfo
  {author} {\bibfnamefont {K.}~\bibnamefont {O'Brien}}, \bibinfo {author}
  {\bibfnamefont {J.~M.}\ \bibnamefont {Kreikebaum}}, \bibinfo {author}
  {\bibfnamefont {D.}~\bibnamefont {Dahlen}}, \bibinfo {author} {\bibfnamefont
  {A.}~\bibnamefont {Morvan}}, \bibinfo {author} {\bibfnamefont
  {B.}~\bibnamefont {Yoshida}}, \bibinfo {author} {\bibfnamefont {N.~Y.}\
  \bibnamefont {Yao}},\ and\ \bibinfo {author} {\bibfnamefont {I.}~\bibnamefont
  {Siddiqi}},\ }\href {https://doi.org/10.1103/PhysRevX.11.021010} {\bibfield
  {journal} {\bibinfo  {journal} {Phys. Rev. X}\ }\textbf {\bibinfo {volume}
  {11}},\ \bibinfo {pages} {021010} (\bibinfo {year} {2021})}\BibitemShut
  {NoStop}%
\bibitem [{\citenamefont {Chiew}\ \emph {et~al.}(2024)\citenamefont {Chiew},
  \citenamefont {Poirier}, \citenamefont {Mishra}, \citenamefont {Bornheimer},
  \citenamefont {Munro}, \citenamefont {Foon}, \citenamefont {Chen},
  \citenamefont {Lim},\ and\ \citenamefont {Nga}}]{chiew2023scalarization}%
  \BibitemOpen
  \bibfield  {author} {\bibinfo {author} {\bibfnamefont {S.-H.}\ \bibnamefont
  {Chiew}}, \bibinfo {author} {\bibfnamefont {K.}~\bibnamefont {Poirier}},
  \bibinfo {author} {\bibfnamefont {R.}~\bibnamefont {Mishra}}, \bibinfo
  {author} {\bibfnamefont {U.}~\bibnamefont {Bornheimer}}, \bibinfo {author}
  {\bibfnamefont {E.}~\bibnamefont {Munro}}, \bibinfo {author} {\bibfnamefont
  {S.~H.}\ \bibnamefont {Foon}}, \bibinfo {author} {\bibfnamefont {C.~W.}\
  \bibnamefont {Chen}}, \bibinfo {author} {\bibfnamefont {W.~S.}\ \bibnamefont
  {Lim}},\ and\ \bibinfo {author} {\bibfnamefont {C.~W.}\ \bibnamefont {Nga}},\
  }\href {https://doi.org/10.1109/TQE.2024.3386753} {\bibfield  {journal}
  {\bibinfo  {journal} {IEEE Transactions on Quantum Engineering}\ }\textbf
  {\bibinfo {volume} {5}},\ \bibinfo {pages} {1} (\bibinfo {year}
  {2024})}\BibitemShut {NoStop}%
\bibitem [{\citenamefont {Díez-Valle}\ \emph {et~al.}(2023)\citenamefont
  {Díez-Valle}, \citenamefont {Luis-Hita}, \citenamefont {Hernández-Santana},
  \citenamefont {Martínez-García}, \citenamefont {Álvaro Díaz-Fernández},
  \citenamefont {Andrés}, \citenamefont {García-Ripoll}, \citenamefont
  {Sánchez-Martínez},\ and\ \citenamefont
  {Porras}}]{diezvalle2023multiobj-constraints}%
  \BibitemOpen
  \bibfield  {author} {\bibinfo {author} {\bibfnamefont {P.}~\bibnamefont
  {Díez-Valle}}, \bibinfo {author} {\bibfnamefont {J.}~\bibnamefont
  {Luis-Hita}}, \bibinfo {author} {\bibfnamefont {S.}~\bibnamefont
  {Hernández-Santana}}, \bibinfo {author} {\bibfnamefont {F.}~\bibnamefont
  {Martínez-García}}, \bibinfo {author} {\bibnamefont {Álvaro
  Díaz-Fernández}}, \bibinfo {author} {\bibfnamefont {E.}~\bibnamefont
  {Andrés}}, \bibinfo {author} {\bibfnamefont {J.~J.}\ \bibnamefont
  {García-Ripoll}}, \bibinfo {author} {\bibfnamefont {E.}~\bibnamefont
  {Sánchez-Martínez}},\ and\ \bibinfo {author} {\bibfnamefont
  {D.}~\bibnamefont {Porras}},\ }\href
  {https://doi.org/10.1088/2058-9565/ace474} {\bibfield  {journal} {\bibinfo
  {journal} {Quantum Science and Technology}\ }\textbf {\bibinfo {volume}
  {8}},\ \bibinfo {pages} {045009} (\bibinfo {year} {2023})}\BibitemShut
  {NoStop}%
\bibitem [{\citenamefont {Chivilikhin}\ \emph {et~al.}(2020)\citenamefont
  {Chivilikhin}, \citenamefont {Samarin}, \citenamefont {Ulyantsev},
  \citenamefont {Iorsh}, \citenamefont {Oganov},\ and\ \citenamefont
  {Kyriienko}}]{mog_vqe2020}%
  \BibitemOpen
  \bibfield  {author} {\bibinfo {author} {\bibfnamefont {D.}~\bibnamefont
  {Chivilikhin}}, \bibinfo {author} {\bibfnamefont {A.}~\bibnamefont
  {Samarin}}, \bibinfo {author} {\bibfnamefont {V.}~\bibnamefont {Ulyantsev}},
  \bibinfo {author} {\bibfnamefont {I.}~\bibnamefont {Iorsh}}, \bibinfo
  {author} {\bibfnamefont {A.}~\bibnamefont {Oganov}},\ and\ \bibinfo {author}
  {\bibfnamefont {O.}~\bibnamefont {Kyriienko}},\ }\bibfield  {journal}
  {\bibinfo  {journal} {arXiv:2007.04424}\ }\href
  {https://doi.org/10.48550/arXiv.2007.04424} {10.48550/arXiv.2007.04424}
  (\bibinfo {year} {2020})\BibitemShut {NoStop}%
\bibitem [{\citenamefont {Zitzler}\ \emph {et~al.}(2003)\citenamefont
  {Zitzler}, \citenamefont {Thiele}, \citenamefont {Laumanns}, \citenamefont
  {Fonseca},\ and\ \citenamefont {da~Fonseca}}]{ZitzlerTLFF03}%
  \BibitemOpen
  \bibfield  {author} {\bibinfo {author} {\bibfnamefont {E.}~\bibnamefont
  {Zitzler}}, \bibinfo {author} {\bibfnamefont {L.}~\bibnamefont {Thiele}},
  \bibinfo {author} {\bibfnamefont {M.}~\bibnamefont {Laumanns}}, \bibinfo
  {author} {\bibfnamefont {C.~M.}\ \bibnamefont {Fonseca}},\ and\ \bibinfo
  {author} {\bibfnamefont {V.~G.}\ \bibnamefont {da~Fonseca}},\ }\href
  {https://doi.org/10.1109/TEVC.2003.810758} {\bibfield  {journal} {\bibinfo
  {journal} {{IEEE} Trans. Evol. Comput.}\ }\textbf {\bibinfo {volume} {7}},\
  \bibinfo {pages} {117} (\bibinfo {year} {2003})}\BibitemShut {NoStop}%
\bibitem [{\citenamefont {Coello}\ \emph {et~al.}(2007)\citenamefont {Coello},
  \citenamefont {Lamont},\ and\ \citenamefont {van
  Veldhuizen}}]{coello2007evolutionary}%
  \BibitemOpen
  \bibfield  {author} {\bibinfo {author} {\bibfnamefont {C.}~\bibnamefont
  {Coello}}, \bibinfo {author} {\bibfnamefont {G.}~\bibnamefont {Lamont}},\
  and\ \bibinfo {author} {\bibfnamefont {D.}~\bibnamefont {van Veldhuizen}},\
  }\href {https://books.google.de/books?id=2murCij_wHcC} {\emph {\bibinfo
  {title} {Evolutionary Algorithms for Solving Multi-Objective Problems}}},\
  Genetic and Evolutionary Computation\ (\bibinfo  {publisher} {Springer US},\
  \bibinfo {year} {2007})\BibitemShut {NoStop}%
\bibitem [{\citenamefont {Beume}\ \emph {et~al.}(2009)\citenamefont {Beume},
  \citenamefont {Fonseca}, \citenamefont {L{\'{o}}pez{-}Ib{\'{a}}{\~{n}}ez},
  \citenamefont {Paquete},\ and\ \citenamefont {Vahrenhold}}]{BeumeFLPV09}%
  \BibitemOpen
  \bibfield  {author} {\bibinfo {author} {\bibfnamefont {N.}~\bibnamefont
  {Beume}}, \bibinfo {author} {\bibfnamefont {C.~M.}\ \bibnamefont {Fonseca}},
  \bibinfo {author} {\bibfnamefont {M.}~\bibnamefont
  {L{\'{o}}pez{-}Ib{\'{a}}{\~{n}}ez}}, \bibinfo {author} {\bibfnamefont
  {L.}~\bibnamefont {Paquete}},\ and\ \bibinfo {author} {\bibfnamefont
  {J.}~\bibnamefont {Vahrenhold}},\ }\href
  {https://doi.org/10.1109/TEVC.2009.2015575} {\bibfield  {journal} {\bibinfo
  {journal} {{IEEE} Trans. Evol. Comput.}\ }\textbf {\bibinfo {volume} {13}},\
  \bibinfo {pages} {1075} (\bibinfo {year} {2009})}\BibitemShut {NoStop}%
\bibitem [{\citenamefont {Beume}(2009)}]{Beume09}%
  \BibitemOpen
  \bibfield  {author} {\bibinfo {author} {\bibfnamefont {N.}~\bibnamefont
  {Beume}},\ }\href {https://doi.org/10.1162/EVCO.2009.17.4.17402} {\bibfield
  {journal} {\bibinfo  {journal} {Evol. Comput.}\ }\textbf {\bibinfo {volume}
  {17}},\ \bibinfo {pages} {477} (\bibinfo {year} {2009})}\BibitemShut
  {NoStop}%
\bibitem [{\citenamefont {Guerreiro}\ \emph {et~al.}(2022)\citenamefont
  {Guerreiro}, \citenamefont {Fonseca},\ and\ \citenamefont
  {Paquete}}]{GuerreiroFP21}%
  \BibitemOpen
  \bibfield  {author} {\bibinfo {author} {\bibfnamefont {A.~P.}\ \bibnamefont
  {Guerreiro}}, \bibinfo {author} {\bibfnamefont {C.~M.}\ \bibnamefont
  {Fonseca}},\ and\ \bibinfo {author} {\bibfnamefont {L.}~\bibnamefont
  {Paquete}},\ }\href {https://doi.org/10.1145/3453474} {\bibfield  {journal}
  {\bibinfo  {journal} {{ACM} Comput. Surv.}\ }\textbf {\bibinfo {volume}
  {54}},\ \bibinfo {pages} {119:1} (\bibinfo {year} {2022})}\BibitemShut
  {NoStop}%
\bibitem [{\citenamefont {Zitzler}\ \emph {et~al.}(2007)\citenamefont
  {Zitzler}, \citenamefont {Brockhoff},\ and\ \citenamefont
  {Thiele}}]{ZitzlerBT06}%
  \BibitemOpen
  \bibfield  {author} {\bibinfo {author} {\bibfnamefont {E.}~\bibnamefont
  {Zitzler}}, \bibinfo {author} {\bibfnamefont {D.}~\bibnamefont {Brockhoff}},\
  and\ \bibinfo {author} {\bibfnamefont {L.}~\bibnamefont {Thiele}},\ }in\
  \href {https://doi.org/10.1007/978-3-540-70928-2\_64} {\emph {\bibinfo
  {booktitle} {Evolutionary Multi-Criterion Optimization, 4th International
  Conference, {EMO} 2007, Matsushima, Japan, March 5-8, 2007, Proceedings}}},\
  \bibinfo {series} {Lecture Notes in Computer Science}, Vol.\ \bibinfo
  {volume} {4403},\ \bibinfo {editor} {edited by\ \bibinfo {editor}
  {\bibfnamefont {S.}~\bibnamefont {Obayashi}}, \bibinfo {editor}
  {\bibfnamefont {K.}~\bibnamefont {Deb}}, \bibinfo {editor} {\bibfnamefont
  {C.}~\bibnamefont {Poloni}}, \bibinfo {editor} {\bibfnamefont
  {T.}~\bibnamefont {Hiroyasu}},\ and\ \bibinfo {editor} {\bibfnamefont
  {T.}~\bibnamefont {Murata}}}\ (\bibinfo  {publisher} {Springer},\ \bibinfo
  {year} {2007})\ pp.\ \bibinfo {pages} {862--876}\BibitemShut {NoStop}%
\bibitem [{\citenamefont {Falc{\'{o}}n{-}Cardona}\ \emph
  {et~al.}(2021)\citenamefont {Falc{\'{o}}n{-}Cardona}, \citenamefont
  {Mart{\'{\i}}nez},\ and\ \citenamefont
  {Garc{\'{\i}}a{-}N{\'{a}}jera}}]{Falcon-CardonaM21}%
  \BibitemOpen
  \bibfield  {author} {\bibinfo {author} {\bibfnamefont {J.~G.}\ \bibnamefont
  {Falc{\'{o}}n{-}Cardona}}, \bibinfo {author} {\bibfnamefont {S.~Z.}\
  \bibnamefont {Mart{\'{\i}}nez}},\ and\ \bibinfo {author} {\bibfnamefont
  {A.}~\bibnamefont {Garc{\'{\i}}a{-}N{\'{a}}jera}},\ }in\ \href
  {https://doi.org/10.1145/3449639.3459276} {\emph {\bibinfo {booktitle}
  {{GECCO} '21: Genetic and Evolutionary Computation Conference, Lille, France,
  July 10-14, 2021}}},\ \bibinfo {editor} {edited by\ \bibinfo {editor}
  {\bibfnamefont {F.}~\bibnamefont {Chicano}}\ and\ \bibinfo {editor}
  {\bibfnamefont {K.}~\bibnamefont {Krawiec}}}\ (\bibinfo  {publisher}
  {{ACM}},\ \bibinfo {year} {2021})\ pp.\ \bibinfo {pages}
  {395--402}\BibitemShut {NoStop}%
\bibitem [{\citenamefont {Falc{\'{o}}n{-}Cardona}\ \emph
  {et~al.}(2022)\citenamefont {Falc{\'{o}}n{-}Cardona}, \citenamefont
  {Emmerich},\ and\ \citenamefont {Coello}}]{Falcon-CardonaE22}%
  \BibitemOpen
  \bibfield  {author} {\bibinfo {author} {\bibfnamefont {J.~G.}\ \bibnamefont
  {Falc{\'{o}}n{-}Cardona}}, \bibinfo {author} {\bibfnamefont {M.~T.~M.}\
  \bibnamefont {Emmerich}},\ and\ \bibinfo {author} {\bibfnamefont {C.~A.~C.}\
  \bibnamefont {Coello}},\ }\href {https://doi.org/10.1162/EVCO\_A\_00307}
  {\bibfield  {journal} {\bibinfo  {journal} {Evol. Comput.}\ }\textbf
  {\bibinfo {volume} {30}},\ \bibinfo {pages} {381} (\bibinfo {year}
  {2022})}\BibitemShut {NoStop}%
\bibitem [{\citenamefont {Auger}\ \emph {et~al.}(2009)\citenamefont {Auger},
  \citenamefont {Bader}, \citenamefont {Brockhoff},\ and\ \citenamefont
  {Zitzler}}]{auger2009hypervolume}%
  \BibitemOpen
  \bibfield  {author} {\bibinfo {author} {\bibfnamefont {A.}~\bibnamefont
  {Auger}}, \bibinfo {author} {\bibfnamefont {J.}~\bibnamefont {Bader}},
  \bibinfo {author} {\bibfnamefont {D.}~\bibnamefont {Brockhoff}},\ and\
  \bibinfo {author} {\bibfnamefont {E.}~\bibnamefont {Zitzler}},\ }in\ \href
  {https://doi.org/10.1145/1527125.1527138} {\emph {\bibinfo {booktitle}
  {Proceedings of the Tenth ACM SIGEVO Workshop on Foundations of Genetic
  Algorithms}}},\ \bibinfo {series and number} {FOGA '09}\ (\bibinfo
  {publisher} {Association for Computing Machinery},\ \bibinfo {address} {New
  York, NY, USA},\ \bibinfo {year} {2009})\ p.\ \bibinfo {pages}
  {87–102}\BibitemShut {NoStop}%
\bibitem [{\citenamefont {Kasper}\ \emph {et~al.}(2022)\citenamefont {Kasper},
  \citenamefont {González-Cuadra}, \citenamefont {Hegde}, \citenamefont {Xia},
  \citenamefont {Dauphin}, \citenamefont {Huber}, \citenamefont {Tiemann},
  \citenamefont {Lewenstein}, \citenamefont {Jendrzejewski},\ and\
  \citenamefont {Hauke}}]{kasper_universal_2022}%
  \BibitemOpen
  \bibfield  {author} {\bibinfo {author} {\bibfnamefont {V.}~\bibnamefont
  {Kasper}}, \bibinfo {author} {\bibfnamefont {D.}~\bibnamefont
  {González-Cuadra}}, \bibinfo {author} {\bibfnamefont {A.}~\bibnamefont
  {Hegde}}, \bibinfo {author} {\bibfnamefont {A.}~\bibnamefont {Xia}}, \bibinfo
  {author} {\bibfnamefont {A.}~\bibnamefont {Dauphin}}, \bibinfo {author}
  {\bibfnamefont {F.}~\bibnamefont {Huber}}, \bibinfo {author} {\bibfnamefont
  {E.}~\bibnamefont {Tiemann}}, \bibinfo {author} {\bibfnamefont
  {M.}~\bibnamefont {Lewenstein}}, \bibinfo {author} {\bibfnamefont
  {F.}~\bibnamefont {Jendrzejewski}},\ and\ \bibinfo {author} {\bibfnamefont
  {P.}~\bibnamefont {Hauke}},\ }\href
  {https://doi.org/10.1088/2058-9565/ac2d39} {\bibfield  {journal} {\bibinfo
  {journal} {Quantum Science and Technology}\ }\textbf {\bibinfo {volume}
  {7}},\ \bibinfo {pages} {015008} (\bibinfo {year} {2022})}\BibitemShut
  {NoStop}%
\bibitem [{\citenamefont {Giorda}\ \emph {et~al.}(2003)\citenamefont {Giorda},
  \citenamefont {Zanardi},\ and\ \citenamefont
  {Lloyd}}]{giorda_universal_2003}%
  \BibitemOpen
  \bibfield  {author} {\bibinfo {author} {\bibfnamefont {P.}~\bibnamefont
  {Giorda}}, \bibinfo {author} {\bibfnamefont {P.}~\bibnamefont {Zanardi}},\
  and\ \bibinfo {author} {\bibfnamefont {S.}~\bibnamefont {Lloyd}},\ }\href
  {https://doi.org/10.1103/PhysRevA.68.062320} {\bibfield  {journal} {\bibinfo
  {journal} {Physical Review A}\ }\textbf {\bibinfo {volume} {68}},\ \bibinfo
  {pages} {062320} (\bibinfo {year} {2003})}\BibitemShut {NoStop}%
\bibitem [{\citenamefont {Barnett}\ and\ \citenamefont
  {Croke}(2009)}]{barnett_quantum_2009}%
  \BibitemOpen
  \bibfield  {author} {\bibinfo {author} {\bibfnamefont {S.~M.}\ \bibnamefont
  {Barnett}}\ and\ \bibinfo {author} {\bibfnamefont {S.}~\bibnamefont
  {Croke}},\ }\href {https://doi.org/10.1364/AOP.1.000238} {\bibfield
  {journal} {\bibinfo  {journal} {Advances in Optics and Photonics}\ }\textbf
  {\bibinfo {volume} {1}},\ \bibinfo {pages} {238} (\bibinfo {year}
  {2009})}\BibitemShut {NoStop}%
\bibitem [{\citenamefont {Kurmapu}\ \emph {et~al.}(2023)\citenamefont
  {Kurmapu}, \citenamefont {Tiunova}, \citenamefont {Tiunov}, \citenamefont
  {Ringbauer}, \citenamefont {Maier}, \citenamefont {Blatt}, \citenamefont
  {Monz}, \citenamefont {Fedorov},\ and\ \citenamefont
  {Lvovsky}}]{kurmapuTomography2023}%
  \BibitemOpen
  \bibfield  {author} {\bibinfo {author} {\bibfnamefont {M.~K.}\ \bibnamefont
  {Kurmapu}}, \bibinfo {author} {\bibfnamefont {V.}~\bibnamefont {Tiunova}},
  \bibinfo {author} {\bibfnamefont {E.}~\bibnamefont {Tiunov}}, \bibinfo
  {author} {\bibfnamefont {M.}~\bibnamefont {Ringbauer}}, \bibinfo {author}
  {\bibfnamefont {C.}~\bibnamefont {Maier}}, \bibinfo {author} {\bibfnamefont
  {R.}~\bibnamefont {Blatt}}, \bibinfo {author} {\bibfnamefont
  {T.}~\bibnamefont {Monz}}, \bibinfo {author} {\bibfnamefont {A.~K.}\
  \bibnamefont {Fedorov}},\ and\ \bibinfo {author} {\bibfnamefont
  {A.}~\bibnamefont {Lvovsky}},\ }\href
  {https://doi.org/10.1103/PRXQuantum.4.040345} {\bibfield  {journal} {\bibinfo
   {journal} {PRX Quantum}\ }\textbf {\bibinfo {volume} {4}},\ \bibinfo {pages}
  {040345} (\bibinfo {year} {2023})}\BibitemShut {NoStop}%
\bibitem [{\citenamefont {Paris}\ and\ \citenamefont
  {Řeháček}(2004)}]{parisTomography2004}%
  \BibitemOpen
  \bibinfo {editor} {\bibfnamefont {M.}~\bibnamefont {Paris}}\ and\ \bibinfo
  {editor} {\bibfnamefont {J.}~\bibnamefont {Řeháček}},\ eds.,\ \href
  {https://doi.org/https://doi.org/10.1007/b98673} {\emph {\bibinfo {title}
  {Quantum State Estimation}}},\ Lecture Notes in Physics\ (\bibinfo
  {publisher} {Springer Berlin, Heidelberg},\ \bibinfo {year}
  {2004})\BibitemShut {NoStop}%
\bibitem [{\citenamefont {Schmitt}\ and\ \citenamefont {{Ekstr\o
  m}}(2024)}]{qmoo_benchmarks}%
  \BibitemOpen
  \bibfield  {author} {\bibinfo {author} {\bibfnamefont {S.}~\bibnamefont
  {Schmitt}}\ and\ \bibinfo {author} {\bibfnamefont {L.}~\bibnamefont {{Ekstr\o
  m}}},\ }\href@noop {} {\bibinfo {title} {qmoo benchmark functions}},\
  \bibinfo {howpublished}
  {\url{https://github.com/HRI-EU/qmoo_benchmark_functions}} (\bibinfo {year}
  {2024})\BibitemShut {NoStop}%
\bibitem [{\citenamefont {Virtanen}\ \emph {et~al.}(2020)\citenamefont
  {Virtanen}, \citenamefont {Gommers}, \citenamefont {Oliphant}, \citenamefont
  {Haberland}, \citenamefont {Reddy}, \citenamefont {Cournapeau}, \citenamefont
  {Burovski}, \citenamefont {Peterson}, \citenamefont {Weckesser},
  \citenamefont {Bright}, \citenamefont {{van der Walt}}, \citenamefont
  {Brett}, \citenamefont {Wilson}, \citenamefont {Millman}, \citenamefont
  {Mayorov}, \citenamefont {Nelson}, \citenamefont {Jones}, \citenamefont
  {Kern}, \citenamefont {Larson}, \citenamefont {Carey}, \citenamefont {Polat},
  \citenamefont {Feng}, \citenamefont {Moore}, \citenamefont {{VanderPlas}},
  \citenamefont {Laxalde}, \citenamefont {Perktold}, \citenamefont {Cimrman},
  \citenamefont {Henriksen}, \citenamefont {Quintero}, \citenamefont {Harris},
  \citenamefont {Archibald}, \citenamefont {Ribeiro}, \citenamefont
  {Pedregosa}, \citenamefont {{van Mulbregt}},\ and\ \citenamefont {{SciPy 1.0
  Contributors}}}]{virtanen_scipy_2020}%
  \BibitemOpen
  \bibfield  {author} {\bibinfo {author} {\bibfnamefont {P.}~\bibnamefont
  {Virtanen}}, \bibinfo {author} {\bibfnamefont {R.}~\bibnamefont {Gommers}},
  \bibinfo {author} {\bibfnamefont {T.~E.}\ \bibnamefont {Oliphant}}, \bibinfo
  {author} {\bibfnamefont {M.}~\bibnamefont {Haberland}}, \bibinfo {author}
  {\bibfnamefont {T.}~\bibnamefont {Reddy}}, \bibinfo {author} {\bibfnamefont
  {D.}~\bibnamefont {Cournapeau}}, \bibinfo {author} {\bibfnamefont
  {E.}~\bibnamefont {Burovski}}, \bibinfo {author} {\bibfnamefont
  {P.}~\bibnamefont {Peterson}}, \bibinfo {author} {\bibfnamefont
  {W.}~\bibnamefont {Weckesser}}, \bibinfo {author} {\bibfnamefont
  {J.}~\bibnamefont {Bright}}, \bibinfo {author} {\bibfnamefont {S.~J.}\
  \bibnamefont {{van der Walt}}}, \bibinfo {author} {\bibfnamefont
  {M.}~\bibnamefont {Brett}}, \bibinfo {author} {\bibfnamefont
  {J.}~\bibnamefont {Wilson}}, \bibinfo {author} {\bibfnamefont {K.~J.}\
  \bibnamefont {Millman}}, \bibinfo {author} {\bibfnamefont {N.}~\bibnamefont
  {Mayorov}}, \bibinfo {author} {\bibfnamefont {A.~R.~J.}\ \bibnamefont
  {Nelson}}, \bibinfo {author} {\bibfnamefont {E.}~\bibnamefont {Jones}},
  \bibinfo {author} {\bibfnamefont {R.}~\bibnamefont {Kern}}, \bibinfo {author}
  {\bibfnamefont {E.}~\bibnamefont {Larson}}, \bibinfo {author} {\bibfnamefont
  {C.~J.}\ \bibnamefont {Carey}}, \bibinfo {author} {\bibfnamefont
  {{\.I}.}~\bibnamefont {Polat}}, \bibinfo {author} {\bibfnamefont
  {Y.}~\bibnamefont {Feng}}, \bibinfo {author} {\bibfnamefont {E.~W.}\
  \bibnamefont {Moore}}, \bibinfo {author} {\bibfnamefont {J.}~\bibnamefont
  {{VanderPlas}}}, \bibinfo {author} {\bibfnamefont {D.}~\bibnamefont
  {Laxalde}}, \bibinfo {author} {\bibfnamefont {J.}~\bibnamefont {Perktold}},
  \bibinfo {author} {\bibfnamefont {R.}~\bibnamefont {Cimrman}}, \bibinfo
  {author} {\bibfnamefont {I.}~\bibnamefont {Henriksen}}, \bibinfo {author}
  {\bibfnamefont {E.~A.}\ \bibnamefont {Quintero}}, \bibinfo {author}
  {\bibfnamefont {C.~R.}\ \bibnamefont {Harris}}, \bibinfo {author}
  {\bibfnamefont {A.~M.}\ \bibnamefont {Archibald}}, \bibinfo {author}
  {\bibfnamefont {A.~H.}\ \bibnamefont {Ribeiro}}, \bibinfo {author}
  {\bibfnamefont {F.}~\bibnamefont {Pedregosa}}, \bibinfo {author}
  {\bibfnamefont {P.}~\bibnamefont {{van Mulbregt}}},\ and\ \bibinfo {author}
  {\bibnamefont {{SciPy 1.0 Contributors}}},\ }\href
  {https://doi.org/10.1038/s41592-019-0686-2} {\bibfield  {journal} {\bibinfo
  {journal} {Nature Methods}\ }\textbf {\bibinfo {volume} {17}},\ \bibinfo
  {pages} {261} (\bibinfo {year} {2020})}\BibitemShut {NoStop}%
\bibitem [{\citenamefont {Hansen}(2006)}]{hansenCMA}%
  \BibitemOpen
  \bibfield  {author} {\bibinfo {author} {\bibfnamefont {N.}~\bibnamefont
  {Hansen}},\ }in\ \href {https://doi.org/10.1007/3-540-32494-1_4} {\emph
  {\bibinfo {booktitle} {Towards a New Evolutionary Computation: Advances in
  the Estimation of Distribution Algorithms}}},\ \bibinfo {editor} {edited by\
  \bibinfo {editor} {\bibfnamefont {J.~A.}\ \bibnamefont {Lozano}}, \bibinfo
  {editor} {\bibfnamefont {P.}~\bibnamefont {Larra{\~{n}}aga}}, \bibinfo
  {editor} {\bibfnamefont {I.}~\bibnamefont {Inza}},\ and\ \bibinfo {editor}
  {\bibfnamefont {E.}~\bibnamefont {Bengoetxea}}}\ (\bibinfo  {publisher}
  {Springer Berlin Heidelberg},\ \bibinfo {address} {Berlin, Heidelberg},\
  \bibinfo {year} {2006})\ pp.\ \bibinfo {pages} {75--102}\BibitemShut
  {NoStop}%
\bibitem [{\citenamefont {Shibata}(2022)}]{CMAESRepo}%
  \BibitemOpen
  \bibfield  {author} {\bibinfo {author} {\bibfnamefont {M.}~\bibnamefont
  {Shibata}},\ }\href@noop {} {\bibinfo {title} {{CMA-ES}}},\ \bibinfo
  {howpublished} {\url{https://github.com/CyberAgentAILab/cmaes}} (\bibinfo
  {year} {2022})\BibitemShut {NoStop}%
\bibitem [{\citenamefont {Kennedy}\ and\ \citenamefont
  {Eberhart}(1995)}]{pso1995}%
  \BibitemOpen
  \bibfield  {author} {\bibinfo {author} {\bibfnamefont {J.}~\bibnamefont
  {Kennedy}}\ and\ \bibinfo {author} {\bibfnamefont {R.}~\bibnamefont
  {Eberhart}},\ }in\ \href {https://doi.org/10.1109/ICNN.1995.488968} {\emph
  {\bibinfo {booktitle} {Proceedings of ICNN'95 - International Conference on
  Neural Networks}}},\ Vol.~\bibinfo {volume} {4}\ (\bibinfo {year} {1995})\
  pp.\ \bibinfo {pages} {1942--1948 vol.4}\BibitemShut {NoStop}%
\bibitem [{\citenamefont {Storn}\ and\ \citenamefont {Price}(1997)}]{de_2997}%
  \BibitemOpen
  \bibfield  {author} {\bibinfo {author} {\bibfnamefont {R.}~\bibnamefont
  {Storn}}\ and\ \bibinfo {author} {\bibfnamefont {K.}~\bibnamefont {Price}},\
  }\href {https://doi.org/https://doi.org/10.1023/A:1008202821328} {\bibfield
  {journal} {\bibinfo  {journal} {Journal of Global Optimization}\ }\textbf
  {\bibinfo {volume} {11}},\ \bibinfo {pages} {341–359} (\bibinfo {year}
  {1997})}\BibitemShut {NoStop}%
\bibitem [{\citenamefont {Hadka}(2022)}]{platypus}%
  \BibitemOpen
  \bibfield  {author} {\bibinfo {author} {\bibfnamefont {D.}~\bibnamefont
  {Hadka}},\ }\href@noop {} {\bibinfo {title} {{Platypus - Multiobjective
  Optimization in Python, v1.1.0}}},\ \bibinfo {howpublished}
  {\url{https://github.com/Project-Platypus/Platypus}} (\bibinfo {year}
  {2022})\BibitemShut {NoStop}%
\bibitem [{\citenamefont {Wolpert}(1996)}]{wolpertNoFreeLunchLearning96}%
  \BibitemOpen
  \bibfield  {author} {\bibinfo {author} {\bibfnamefont {D.~H.}\ \bibnamefont
  {Wolpert}},\ }\href {https://doi.org/10.1162/neco.1996.8.7.1341} {\bibfield
  {journal} {\bibinfo  {journal} {Neural Computation}\ }\textbf {\bibinfo
  {volume} {8}},\ \bibinfo {pages} {1341} (\bibinfo {year} {1996})},\ \Eprint
  {https://arxiv.org/abs/https://direct.mit.edu/neco/article-pdf/8/7/1341/813495/neco.1996.8.7.1341.pdf}
  {https://direct.mit.edu/neco/article-pdf/8/7/1341/813495/neco.1996.8.7.1341.pdf}
  \BibitemShut {NoStop}%
\bibitem [{\citenamefont {Wolpert}\ and\ \citenamefont
  {Macready}(1997)}]{wolpertNoFreeLunchOpt97}%
  \BibitemOpen
  \bibfield  {author} {\bibinfo {author} {\bibfnamefont {D.}~\bibnamefont
  {Wolpert}}\ and\ \bibinfo {author} {\bibfnamefont {W.}~\bibnamefont
  {Macready}},\ }\href {https://doi.org/10.1109/4235.585893} {\bibfield
  {journal} {\bibinfo  {journal} {IEEE Transactions on Evolutionary
  Computation}\ }\textbf {\bibinfo {volume} {1}},\ \bibinfo {pages} {67}
  (\bibinfo {year} {1997})}\BibitemShut {NoStop}%
\bibitem [{\citenamefont {van Rijn}\ \emph {et~al.}(2017)\citenamefont {van
  Rijn}, \citenamefont {Wang}, \citenamefont {van Stein},\ and\ \citenamefont
  {B\"{a}ck}}]{sanderAlgo2017}%
  \BibitemOpen
  \bibfield  {author} {\bibinfo {author} {\bibfnamefont {S.}~\bibnamefont {van
  Rijn}}, \bibinfo {author} {\bibfnamefont {H.}~\bibnamefont {Wang}}, \bibinfo
  {author} {\bibfnamefont {B.}~\bibnamefont {van Stein}},\ and\ \bibinfo
  {author} {\bibfnamefont {T.}~\bibnamefont {B\"{a}ck}},\ }in\ \href
  {https://doi.org/10.1145/3071178.3071205} {\emph {\bibinfo {booktitle}
  {Proceedings of the Genetic and Evolutionary Computation Conference}}},\
  \bibinfo {series and number} {GECCO '17}\ (\bibinfo  {publisher} {Association
  for Computing Machinery},\ \bibinfo {address} {New York, NY, USA},\ \bibinfo
  {year} {2017})\ p.\ \bibinfo {pages} {737–744}\BibitemShut {NoStop}%
\bibitem [{\citenamefont {López-Ibáñez}\ \emph {et~al.}(2016)\citenamefont
  {López-Ibáñez}, \citenamefont {Dubois-Lacoste}, \citenamefont {{Pérez
  Cáceres}}, \citenamefont {Birattari},\ and\ \citenamefont
  {Stützle}}]{stuetzleIRACE2016}%
  \BibitemOpen
  \bibfield  {author} {\bibinfo {author} {\bibfnamefont {M.}~\bibnamefont
  {López-Ibáñez}}, \bibinfo {author} {\bibfnamefont {J.}~\bibnamefont
  {Dubois-Lacoste}}, \bibinfo {author} {\bibfnamefont {L.}~\bibnamefont
  {{Pérez Cáceres}}}, \bibinfo {author} {\bibfnamefont {M.}~\bibnamefont
  {Birattari}},\ and\ \bibinfo {author} {\bibfnamefont {T.}~\bibnamefont
  {Stützle}},\ }\href
  {https://doi.org/https://doi.org/10.1016/j.orp.2016.09.002} {\bibfield
  {journal} {\bibinfo  {journal} {Operations Research Perspectives}\ }\textbf
  {\bibinfo {volume} {3}},\ \bibinfo {pages} {43} (\bibinfo {year}
  {2016})}\BibitemShut {NoStop}%
\bibitem [{\citenamefont {Egger}\ \emph {et~al.}(2021)\citenamefont {Egger},
  \citenamefont {Mare{\v{c}}ek},\ and\ \citenamefont
  {Woerner}}]{Egger2021warmstartingquantum}%
  \BibitemOpen
  \bibfield  {author} {\bibinfo {author} {\bibfnamefont {D.~J.}\ \bibnamefont
  {Egger}}, \bibinfo {author} {\bibfnamefont {J.}~\bibnamefont
  {Mare{\v{c}}ek}},\ and\ \bibinfo {author} {\bibfnamefont {S.}~\bibnamefont
  {Woerner}},\ }\href {https://doi.org/10.22331/q-2021-06-17-479} {\bibfield
  {journal} {\bibinfo  {journal} {{Quantum}}\ }\textbf {\bibinfo {volume}
  {5}},\ \bibinfo {pages} {479} (\bibinfo {year} {2021})}\BibitemShut {NoStop}%
\bibitem [{\citenamefont {Akshay}\ \emph {et~al.}(2021)\citenamefont {Akshay},
  \citenamefont {Rabinovich}, \citenamefont {Campos},\ and\ \citenamefont
  {Biamonte}}]{akshayParameterConQAOA2021}%
  \BibitemOpen
  \bibfield  {author} {\bibinfo {author} {\bibfnamefont {V.}~\bibnamefont
  {Akshay}}, \bibinfo {author} {\bibfnamefont {D.}~\bibnamefont {Rabinovich}},
  \bibinfo {author} {\bibfnamefont {E.}~\bibnamefont {Campos}},\ and\ \bibinfo
  {author} {\bibfnamefont {J.}~\bibnamefont {Biamonte}},\ }\href
  {https://doi.org/10.1103/PhysRevA.104.L010401} {\bibfield  {journal}
  {\bibinfo  {journal} {Phys. Rev. A}\ }\textbf {\bibinfo {volume} {104}},\
  \bibinfo {pages} {L010401} (\bibinfo {year} {2021})}\BibitemShut {NoStop}%
\bibitem [{\citenamefont {Sack}\ \emph {et~al.}(2023)\citenamefont {Sack},
  \citenamefont {Medina}, \citenamefont {Kueng},\ and\ \citenamefont
  {Serbyn}}]{sackQAOA_init2023}%
  \BibitemOpen
  \bibfield  {author} {\bibinfo {author} {\bibfnamefont {S.~H.}\ \bibnamefont
  {Sack}}, \bibinfo {author} {\bibfnamefont {R.~A.}\ \bibnamefont {Medina}},
  \bibinfo {author} {\bibfnamefont {R.}~\bibnamefont {Kueng}},\ and\ \bibinfo
  {author} {\bibfnamefont {M.}~\bibnamefont {Serbyn}},\ }\href
  {https://doi.org/10.1103/PhysRevA.107.062404} {\bibfield  {journal} {\bibinfo
   {journal} {Phys. Rev. A}\ }\textbf {\bibinfo {volume} {107}},\ \bibinfo
  {pages} {062404} (\bibinfo {year} {2023})}\BibitemShut {NoStop}%
\bibitem [{\citenamefont {Jain}\ \emph {et~al.}(2022)\citenamefont {Jain},
  \citenamefont {Coyle}, \citenamefont {Kashefi},\ and\ \citenamefont
  {Kumar}}]{Jain2022graphneuralnetwork}%
  \BibitemOpen
  \bibfield  {author} {\bibinfo {author} {\bibfnamefont {N.}~\bibnamefont
  {Jain}}, \bibinfo {author} {\bibfnamefont {B.}~\bibnamefont {Coyle}},
  \bibinfo {author} {\bibfnamefont {E.}~\bibnamefont {Kashefi}},\ and\ \bibinfo
  {author} {\bibfnamefont {N.}~\bibnamefont {Kumar}},\ }\href
  {https://doi.org/10.22331/q-2022-11-17-861} {\bibfield  {journal} {\bibinfo
  {journal} {{Quantum}}\ }\textbf {\bibinfo {volume} {6}},\ \bibinfo {pages}
  {861} (\bibinfo {year} {2022})}\BibitemShut {NoStop}%
\bibitem [{\citenamefont {Bravyi}\ \emph {et~al.}(2020)\citenamefont {Bravyi},
  \citenamefont {Kliesch}, \citenamefont {Koenig},\ and\ \citenamefont
  {Tang}}]{bravyiRecursive2020}%
  \BibitemOpen
  \bibfield  {author} {\bibinfo {author} {\bibfnamefont {S.}~\bibnamefont
  {Bravyi}}, \bibinfo {author} {\bibfnamefont {A.}~\bibnamefont {Kliesch}},
  \bibinfo {author} {\bibfnamefont {R.}~\bibnamefont {Koenig}},\ and\ \bibinfo
  {author} {\bibfnamefont {E.}~\bibnamefont {Tang}},\ }\href
  {https://doi.org/10.1103/PhysRevLett.125.260505} {\bibfield  {journal}
  {\bibinfo  {journal} {Phys. Rev. Lett.}\ }\textbf {\bibinfo {volume} {125}},\
  \bibinfo {pages} {260505} (\bibinfo {year} {2020})}\BibitemShut {NoStop}%
\bibitem [{\citenamefont {Scriva}\ \emph {et~al.}(2024)\citenamefont {Scriva},
  \citenamefont {Astrakhantsev}, \citenamefont {Pilati},\ and\ \citenamefont
  {Mazzola}}]{scrivaVaraitionalShots2024}%
  \BibitemOpen
  \bibfield  {author} {\bibinfo {author} {\bibfnamefont {G.}~\bibnamefont
  {Scriva}}, \bibinfo {author} {\bibfnamefont {N.}~\bibnamefont
  {Astrakhantsev}}, \bibinfo {author} {\bibfnamefont {S.}~\bibnamefont
  {Pilati}},\ and\ \bibinfo {author} {\bibfnamefont {G.}~\bibnamefont
  {Mazzola}},\ }\href {https://doi.org/10.1103/PhysRevA.109.032408} {\bibfield
  {journal} {\bibinfo  {journal} {Phys. Rev. A}\ }\textbf {\bibinfo {volume}
  {109}},\ \bibinfo {pages} {032408} (\bibinfo {year} {2024})}\BibitemShut
  {NoStop}%
\bibitem [{\citenamefont {Herrman}\ \emph {et~al.}(2022)\citenamefont
  {Herrman}, \citenamefont {Lotshaw}, \citenamefont {Ostrowski}, \citenamefont
  {Humble},\ and\ \citenamefont {Siopsis}}]{herrmanMA-QAOA_2022}%
  \BibitemOpen
  \bibfield  {author} {\bibinfo {author} {\bibfnamefont {R.}~\bibnamefont
  {Herrman}}, \bibinfo {author} {\bibfnamefont {P.}~\bibnamefont {Lotshaw}},
  \bibinfo {author} {\bibfnamefont {J.}~\bibnamefont {Ostrowski}}, \bibinfo
  {author} {\bibfnamefont {T.}~\bibnamefont {Humble}},\ and\ \bibinfo {author}
  {\bibfnamefont {G.}~\bibnamefont {Siopsis}},\ }\href
  {https://doi.org/10.1038/s41598-022-10555-8} {\bibfield  {journal} {\bibinfo
  {journal} {Sci Rep}\ }\textbf {\bibinfo {volume} {12}},\ \bibinfo {pages}
  {6781} (\bibinfo {year} {2022})}\BibitemShut {NoStop}%
\bibitem [{\citenamefont {Saleem}\ \emph {et~al.}(2021)\citenamefont {Saleem},
  \citenamefont {Tomesh}, \citenamefont {Tariq},\ and\ \citenamefont
  {Suchara}}]{saleem_localMixingQAOA2021}%
  \BibitemOpen
  \bibfield  {author} {\bibinfo {author} {\bibfnamefont {Z.~H.}\ \bibnamefont
  {Saleem}}, \bibinfo {author} {\bibfnamefont {T.}~\bibnamefont {Tomesh}},
  \bibinfo {author} {\bibfnamefont {B.}~\bibnamefont {Tariq}},\ and\ \bibinfo
  {author} {\bibfnamefont {M.}~\bibnamefont {Suchara}},\ }\href@noop {}
  {\bibfield  {journal} {\bibinfo  {journal} {arXiv:2010.06660}\ } (\bibinfo
  {year} {2021})}\BibitemShut {NoStop}%
\bibitem [{\citenamefont {Shi}\ \emph {et~al.}(2022)\citenamefont {Shi},
  \citenamefont {Herrman}, \citenamefont {Shaydulin}, \citenamefont
  {Chakrabarti}, \citenamefont {Pistoia},\ and\ \citenamefont
  {Larson}}]{shi_localMixerQAOA_2022}%
  \BibitemOpen
  \bibfield  {author} {\bibinfo {author} {\bibfnamefont {K.}~\bibnamefont
  {Shi}}, \bibinfo {author} {\bibfnamefont {R.}~\bibnamefont {Herrman}},
  \bibinfo {author} {\bibfnamefont {R.}~\bibnamefont {Shaydulin}}, \bibinfo
  {author} {\bibfnamefont {S.}~\bibnamefont {Chakrabarti}}, \bibinfo {author}
  {\bibfnamefont {M.}~\bibnamefont {Pistoia}},\ and\ \bibinfo {author}
  {\bibfnamefont {J.}~\bibnamefont {Larson}},\ }in\ \href
  {https://doi.org/10.1109/SEC54971.2022.00062} {\emph {\bibinfo {booktitle}
  {2022 IEEE/ACM 7th Symposium on Edge Computing (SEC)}}}\ (\bibinfo {year}
  {2022})\ pp.\ \bibinfo {pages} {414--419}\BibitemShut {NoStop}%
\bibitem [{\citenamefont {Larocca}\ \emph {et~al.}(2024)\citenamefont
  {Larocca}, \citenamefont {Thanasilp}, \citenamefont {Wang}, \citenamefont
  {Sharma}, \citenamefont {Biamonte}, \citenamefont {Coles}, \citenamefont
  {Cincio}, \citenamefont {McClean}, \citenamefont {Holmes},\ and\
  \citenamefont {Cerezo}}]{laroccaBarrenPlateauReview2024}%
  \BibitemOpen
  \bibfield  {author} {\bibinfo {author} {\bibfnamefont {M.}~\bibnamefont
  {Larocca}}, \bibinfo {author} {\bibfnamefont {S.}~\bibnamefont {Thanasilp}},
  \bibinfo {author} {\bibfnamefont {S.}~\bibnamefont {Wang}}, \bibinfo {author}
  {\bibfnamefont {K.}~\bibnamefont {Sharma}}, \bibinfo {author} {\bibfnamefont
  {J.}~\bibnamefont {Biamonte}}, \bibinfo {author} {\bibfnamefont {P.~J.}\
  \bibnamefont {Coles}}, \bibinfo {author} {\bibfnamefont {L.}~\bibnamefont
  {Cincio}}, \bibinfo {author} {\bibfnamefont {J.~R.}\ \bibnamefont {McClean}},
  \bibinfo {author} {\bibfnamefont {Z.}~\bibnamefont {Holmes}},\ and\ \bibinfo
  {author} {\bibfnamefont {M.}~\bibnamefont {Cerezo}},\ }\bibfield  {journal}
  {\bibinfo  {journal} {arXiv:2405.00781}\ }\href
  {https://doi.org/10.48550/arXiv.2405.00781} {10.48550/arXiv.2405.00781}
  (\bibinfo {year} {2024})\BibitemShut {NoStop}%
\end{thebibliography}
%

\end{document}